\pdfoutput=1
\documentclass[aps, twocolumn, floatfix, superscriptaddress, prb]{revtex4-1}
\usepackage{graphicx}
\DeclareGraphicsExtensions{.pdf}
\usepackage{amsmath,amssymb,bbold,bm,color}
\usepackage{float}
\usepackage{epstopdf}
\usepackage{tabularx}
\usepackage{braket}
\usepackage{booktabs}
\usepackage{array}
\usepackage{blindtext}
\usepackage{natbib}

\usepackage[colorlinks=true,breaklinks=true,linkcolor=blue,anchorcolor=red,citecolor=blue,urlcolor=blue]{hyperref}

\begin{document}

\title{Inner non-Hermitian skin effect on Bethe lattice}

\author{Junsong Sun}
\affiliation{School of Physics, Beihang University,
Beijing, 100191, China}

\author{Chang-An Li}
\email{changan.li@uni-wuerzburg.de}
\affiliation{Institute for Theoretical Physics and Astrophysics, University of W$\ddot{u}$rzburg, 97074 W$\ddot{u}$rzburg, Germany}

\author{Shiping Feng}
\affiliation{ Department of Physics,  Beijing Normal University, Beijing, 100875, China}

\author{Huaiming Guo}
\email{hmguo@buaa.edu.cn}
\affiliation{School of Physics, Beihang University,
Beijing, 100191, China}

\begin{abstract}
We investigate the non-Hermitian Su-Schrieffer-Heeger (SSH) model on Bethe lattice, revealing a novel localization phenomenon coined inner non-Hermitian skin effect. This effect is featured by the localization of all eigenstates within the bulk of the lattice, diverging from the conventional skin effect observed in general non-Hermitian systems. The analytical treatment of the model demonstrates that the Hamiltonian can be decoupled into a series of one-dimensional chains, with one end fixed at the bottom boundary while the other ends positioned at varying generations within the bulk. This configuration leads to the emergence of the inner non-Hermitian skin effect, which is further validated by performing circuit simulations. Our findings provide new insights into the interplay between non-Hermitian physics and the self-similar structure on Bethe lattice. 
\end{abstract}


\maketitle
\textit{Introduction.-}
The Bethe lattice is a tree-like geometry, in which each vertex has the same number of neighbors. Its key feature is the absence of closed loops, which reduces complex quantum interference effects and makes lattice models on the Bethe lattice generally easier to solve compared to other lattices\cite{PhysRevB.63.155110,Aryal_2020}. For instance, simple models like the Ising model can be solved exactly on the Bethe lattice\cite{PhysRevB.34.7975}. Additionally, the study of interesting models on the Bethe lattice is motivated by the similarity of results to those on crystal lattices. Solutions on the Bethe lattice often provide a better approximation than the mean-field approach for thermodynamic quantities of models on regular lattices\cite{PhysRevLett.74.809, PhysRevB.71.235119}, such as square and honeycomb lattices, offering valuable insights into the physical properties of lattice models. Consequently, significant advances have been made in studying various models on the Bethe lattice, including classical statistical models, quantum spin models, spin-glass systems~\cite{PhysRevB.34.7975,PhysRevLett.56.1082, PhysRevE.52.2187, PhysRevB.78.134424,PhysRevB.80.014524, PhysRevB.80.144415,PhysRevB.86.195137,   Daniska2016,PhysRevB.100.125121,PhysRevResearch.3.023054}, the Bose/Fermi Hubbard model, and the Anderson model~\cite{PhysRevB.34.6394,PhysRevLett.72.526,PhysRevLett.113.046806, PhysRevB.100.094201}.


In recent years, there has been a remarkable surge in research interest in non-Hermitian physics\cite{PhysRevLett.121.086803,PhysRevB.97.045106,PhysRevLett.121.136802,GongZ18prx,PhysRevLett.123.246801,LiuT19prl,JinL19prb,Kawabata19prx,PhysRevLett.123.090603,PhysRevLett.122.237601,PhysRevA.100.032102,PhysRevB.99.155431,PhysRevB.101.045422,PhysRevA.101.043811,Zeng20prb,HuH21prl,GuoCX21prl,TianyuLi2021,Bergholtz21rmp,reviewskineffect,PhysRevB.105.094103,PhysRevB.108.245114,PhysRevLett.133.076502,PRXwangzhong,WanerHou2024}, owing to its rich phenomena and promising applications. As the simplest non-Hermitian topological system, the non-Hermitian version of the SSH model has been widely investigated\cite{PhysRevB.97.045106,PhysRevA.100.032102,PhysRevA.101.043811,PhysRevB.99.155431,PhysRevB.101.045422}, revealing many intriguing properties such as the non-Hermitian skin effect and the breakdown of conventional bulk-boundary correspondence\cite{PhysRevLett.121.086803,PhysRevLett.123.246801}. To address these issues, non-Bloch band theory has been constructed in one dimension based on the generalized Brillouin zone, and it has been successfully applied to various non-Hermitian systems\cite{PhysRevLett.123.066404,PhysRevLett.121.136802}. In two and higher dimensions, the skin effect can be manifested in different forms, such as the corner-skin effect and geometry-dependent skin effect\cite{PhysRevLett.122.237601,PhysRevLett.123.016805,PhysRevB.102.205118,PhysRevB.102.241202,FuY21prb,Zhangx21nc,PhysRevLett.128.223903,PhysRevB.106.035425,PhysRevLett.131.116601,PhysRevB.108.075122}. It is proposed that the existence of the skin effect depends on whether the periodic-boundary spectrum of the Hamiltonian covers a finite area in the complex plane\cite{Zhang2022}. For system of arbitrary spatial dimensions,  an amoeba theory for the non-Hermitian skin effect and non-Bloch band theory has also been developed\cite{PRXwangzhong}. For aforementioned Bethe lattice with fractal structures, we have recently shown that the non-Hermiticity can give rise to particular quantum fractals  in terms of energy and wavefunctions on it\cite{Sun24NHQF}, while the application of more general non-Hermitian effects and possible skin effects on Bethe lattice are largely unexplored.

\begin{figure}[hbpt]
  \centering
  \includegraphics[width=8.6cm]{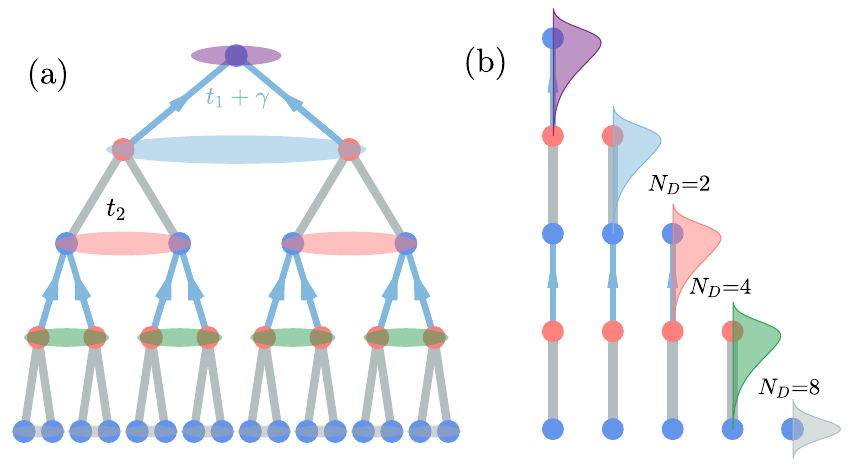}\\
  \caption{(a) Schematic of the non-Hermitian SSH model on the Bethe lattice (or tree lattice) with $G=4$ generations. Blue arrows represent the non-reciprocal hopping amplitude $t_1+\gamma$ (the reverse ones with $t_1-\gamma$ are omitted for clarity). Gray bonds are with an amplitude of $t_2$. The colored connectors denote the inner boundaries, near which the inner skin effect occurs. (b): Decoupled effective 1D chains corresponding to (a) with $N_D$ denoting the degeneracy of states.}\label{fig1}
\end{figure}

In this work, we implement the non-Hermitian SSH model on Bethe lattice and explore its intriguing non-Hermitian features which arise from the interplay between non-Hermitian topology and the self-similar lattice structure. We analytically solve the non-Hermitian SSH model on Bethe lattice by developing a decoupling technique, obtaining the energy spectrum and corresponding eigenstates. We identify a distinct form of non-Hermitian skin effect with eigenstates localizing within the bulk region, coined `inner non-Hermitian skin effect'\cite{innerskin}.  We explain this particular inner non-Hermtian skin effect from the combination of non-Hermiticity and bulk-boundary duality of Bethe lattice. To further show the experimental relevance of our proposal, we implement the non-Hermitian SSH model using a Bethe-lattice circuit, which further validates our theoretical findings.


\textit{Non-Hermitian SSH Model on Bethe lattice.-}
The non-Hermitian SSH model on Bethe lattice is characterized by alternating hopping strengths $t_1\pm\gamma$ and $t_2$ across different generations, as schematically illustrated in Fig. \ref{fig1}. Specifically, the hoppings on the bonds with strength $t_1\pm\gamma$ become non-reciprocal due to the presence of the non-Hermitian term $\gamma$. The Hamiltonian for this model can be expressed as follows:
\begin{align}\label{eq1}
\nonumber
  H=&\sum_{{\langle ij\rangle}_1}\left[(t_1+\gamma)\hat{c}_i\hat{c}_j^\dagger+(t_1-\gamma)\hat{c}_j^\dagger \hat{c}_i\right]\\
  +&\sum_{{\langle ij\rangle}_2}\left[t_2\hat{c}_i\hat{c}_j^\dagger+\text{h.c.}\right],
\end{align}
where $\hat{c}_{i}^{\dagger}$ ($\hat{c}_{i}$) is the creation (annihilation) operator at site $i$; $\langle ij\rangle_{\alpha}$ represents two types of nearest neighbors distinguished by $\alpha=1,2$.
Under the basis $\hat{\Phi}=(\hat{\phi}_0,\hat{\phi}_1,\cdots,\hat{\phi}_G)^T$ with $\hat{\phi}_0=\hat{c}^0_0$ and $\hat{\phi}_l=(\hat{c}^l_1,\hat{c}^l_2,\cdots,\hat{c}^l_{N_l})$ ($N_l=2^{l-1}$ is the number of the sites in the $l$-th generation), the Hamiltonian can be recast as $H=\hat{\Phi}^{\dagger}{\cal H}\hat{\Phi}$.

\begin{figure}[hbpt]
  \centering
  \includegraphics[width=8.8 cm]{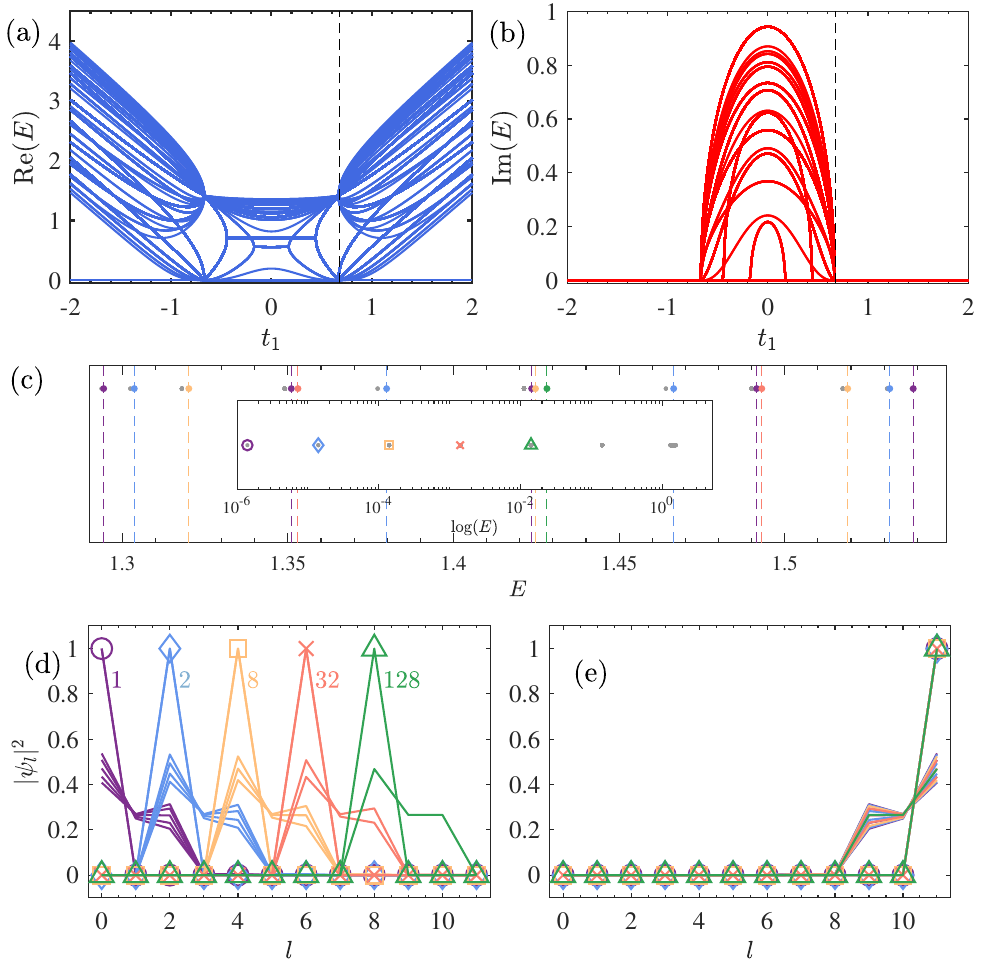}\\
  \caption{The energy spectrum and inner non-Hermitian skin states. The energy spectrum as a function of $t_1$: (a) the real part and (b) the imaginary part. (c) The positive eigenvalues at specific choice of $t_1=0.6741$ $(\bar{t}_1=0.1)$. The negative ones are symmetric and therefore not displayed here. (d) The distribution of eigenstates corresponding to (c) along different generations of the tree lattice. (e) The distribution of the eigenstates at $\gamma=-2/3$, with the other parameters are the same as in (d). The reversal of the sign in $\gamma$ does not affect the eigenvalues, but it change where the skin modes localize. The colors and symbols used in (d) and (e) correspond to those in (c). Here, unless specified otherwise, the parameters $t_2=1$ and $\gamma=2/3$ are used. The tree lattice has $G=11$ generations.}\label{fig2}
\end{figure}

The corresponding Hamiltonian matrix $\mathcal{H}$ possesses a chiral symmetry, defined as $\eta^{-1}\mathcal{H}\eta=-\mathcal{H}$, where $\eta$ is a diagonal matrix given by $\eta=\text{diag}(D_0,-D_1,D_2,\cdots,-D_{G-1},D_G)$, with $D_l$ ($l=0,\cdots, G$) being a $1\times N_l$ matrix with all elements equal to unity. When subjected to this symmetry, the eigenvalues of the Hamiltonian appear in pairs of $(E,-E)$.
Figure \ref{fig2}~(a) illustrates the energy spectrum as a function of $t_1$, while holding $t_2$ and $\gamma$ constant. It is notable that within the range of $|t_1| > |\gamma|$, the energy spectrum remains real and only acquires imaginary components when $|t_1| < |\gamma|$ [see Fig. \ref{fig2}~(b)].

For the range of $t_1$ in which the system shows real energy spectrum only, it is possible to transform the non-Hermitian Hamiltonian in Eq.~(\ref{eq1}) into an equivalent Hermitian one through a similarity transformation $S$, while preserving the eigenvalues of the spectrum~\cite{PhysRevLett.121.086803}.
The operator $S$ can be chosen as a diagonal matrix with the following diagonal elements:
\begin{align}\label{eqS}
\begin{split}
S=\left\{1D_0,rD_1,rD_2,r^2D_3,r^2D_4,\cdots,\right.\\
 \left.r^{\frac{G-1}{2}}D_{G-2},r^{\frac{G-1}{2}}D_{G-1}, r^{\frac{G+1}{2}}D_{G}\right\},
\end{split}
\end{align}
with $r\equiv\sqrt{\frac{\left|t_1-\gamma\right|}{\left|t_1+\gamma\right|}}$.
The resulting Hamiltonian is a Hermitian SSH model $\bar{\mathcal{H}}=S^{-1}\mathcal{H}S$, in which the alternating hopping amplitudes are given by $\bar{t}_1=\sqrt{(t_1-\gamma)(t_1+\gamma)}$ and $\bar{t}_2=t_2$. The wavefunction transforms according to $|\Psi\rangle=S|\bar{\Psi}\rangle$. From the explicit expression of the transformation matrix $S$, it is evident that the presence of the non-Hermitian term $\gamma$ will induce localization behavior in the wavefunctions $|\Psi\rangle$. This localization occurs towards the root site (first generation) for $r<1$ or towards the bottom direction (last generation) for $r>1$.

\textit{Inner non-Hermitian skin effect.-}
The skin effect is an unusual phenomenon in non-Hermitian systems, where all eigenstates tend to localize near boundaries of the system under open boundary conditions. Interestingly, on the Bethe lattice, a different form of non-Hermitian skin effect arises, with eigenstates localizing within the bulk region instead. Explicitly, as illustrated in Fig. \ref{fig2}~(d), various categories of eigenstates localize near different generations within the central region. We refer to such anomalous phenomenon as inner non-Hermitian skin effect\cite{innerskin}.   This behavior shows stark difference from the non-Hermitian skin effect on regular lattices, in which all of the skin modes localize at boundaries. Note that the topologically protected edge states from SSH model emerging at zero energy exhibit similar localization behaviors, but with a higher degree of localization.

The inner non-Hermitian skin effect is unique to Bethe lattice and does not have counterparts in other extensively studied non-Hermitian systems on conventional one-dimensional (1D) or two-dimensional (2D) lattices. It suggests that, in addition to the bottom boundary, the Bethe lattice features inner boundaries that vary in position for different eigenstates of the non-Hermitian SSH model. The appearance of inner non-Hermitian skin effect is related to the non-reciproical hopping directions. When $\gamma > 0$, eigenstates tend to localize towards the root site ($G=1$ generation) and encounter the inner boundaries, resulting in the inner non-Hermitian skin effect. On the contrary, if $\gamma < 0$, the direction of non-reciprocal hopping is inversed, leading wavefunctions to localize near the bottom boundary (last generation sites), resembling the conventional non-Hermitian skin effect.

\textit{Analytic solutions.-}
We then proceed with an analytical solution of the non-Hermitian SSH model on Bethe lattice and explain the appearance of inner non-Hermitian skin effect. Note that the method applied here is suitable for Bethe lattices with arbitrary sub-branching number $z$. We sketch our calculations here and leave full details in Appendix A. To facilitate the calculations, we introduce two different types of states: $\psi_{l}$ and $\phi_{r}^{l,\alpha}$. Here, $\psi_{l}$ denotes a symmetric linear combination of states located on the sites belonging to the $l$-th generation, while $\phi_{r}^{l,\alpha}$ represents a linear combination of wavefunctions at sites on different branches in the $r$-th generation ($r=1,\cdots,G-l$), which are derived from the $\alpha$ site in the $l$-th generation. It should be noted that the coefficients in the linear combination of $\phi_{r}^{l,\alpha}$ satisfy the condition that their sum is zero.
Then the eigenequation $\mathcal{H}|\Psi\rangle=E|\Psi\rangle$ can be decoupled into a set of equations.
The state $\psi_{l}$ satisfies
\begin{align}\label{eqlm}
\begin{split}
  E\psi_{l}=\sqrt{z}[t_{\alpha}\psi_{l-1}+t_{\beta}\psi_{l+1}],
\end{split}
\end{align}
where $t_{\alpha}=t_2~(t_1-\gamma)$ and $t_{\beta}=t_1+\gamma~(t_2)$ for even (odd) values of $l$ with $l=0,1,\cdots, G$, and $\psi_{l}=0$ for $l<0$ or $l>G$ corresponding to open boundary condition.
For $\phi^{l,\alpha}_r$, we have
\begin{align}\label{eqlrlapha}
\begin{split}
  E\phi_{r}^{l,\alpha}&=\sqrt{z}[t_{\alpha}\phi_{r-1}^{l,\alpha}+t_{\beta}\phi_{r+1}^{l,\alpha}],
\end{split}
\end{align}
where the values of $t_{\alpha}, t_{\beta}$ depend on the parity of $l+r$, with $l=0,1,\cdots, G-1$, $r=2,\cdots,G-l$, and $\alpha=1,\cdots,z(z-1)^{l-1}$.
Actually, the above two equations represent the eigenproblem of a 1D non-Hermitian SSH Hamiltonian. As a result, the non-Hermitian SSH on tree lattice can be decoupled into 1D chains with different lengths varying from $1$ to $G+1$, revealing its inherent fractal nature.

Specifically, the set of $G+1$ equations in Eq.~(\ref{eqlm}) represent a 1D non-Hermitian SSH chain with a length of $G+1$.
For $\phi^{l,\alpha}_r(l=0,\cdots,G-1)$, given a specific value of $l$, the possible values of $r$ are determined by $l$~($r=1,\cdots,G-l$). Therefore, the set of equations in Eq.~(\ref{eqlrlapha}) correspond to a series of 1D SSH chains with lengths ranging from $1$ to $G$, each exhibiting a $(z-1)z^l$-fold degeneracy. In total, we can obtain the number of eigenvectors and eigenvalues: $N_{T}=\sum_{l=0}^{G-1}(z-1)z^l(G-l)+G+1$,
which is exactly the total number of sites $\frac{z^{G+1}-1}{z-1}$ in the entire tree lattice.

When $|t_1| > |\gamma|$, the 1D non-Hermitian SSH model can be transformed into a Hermitian one, with the alternating hopping amplitudes being $\bar{t}_1, \bar{t}_2$.
For a general SSH model given by\cite{jps2017}:
\begin{equation}\label{eqSSH}
  \bar{t}_n\varphi_n+\epsilon\varphi_{n+1}+\bar{t}_{n+1}\varphi_{n+2}=0,~~(\bar{t}_{n+2}=\bar{t}_n).
\end{equation}
Under open boundary conditions ($\varphi_0=\varphi_{L+1}=0$), the analytical expression for the eigenvalues can be obtained as
\begin{equation}\label{energySSH}
\epsilon_n=\pm\sqrt{\bar{t}_1^2+\bar{t}_2^2+2\bar{t}_1\bar{t}_2\cos(\alpha_n)},
\end{equation}
with $\alpha_n=\pm\frac{n\pi}{N+1}$ ($n=1,\cdots,N$) for odd length $L=2N+1$, and $\alpha_n$ determined by the boundary condition $\sin[(N+1)\alpha_n]+\frac{\bar{t}_2}{\bar{t}_1}\sin(N\alpha_n)=0$ for even $L=2N$.
Additionally, there exists a zero eigenvalue for the odd-length chains.
Therefore, the eigenvalues of the non-Hermitian SSH model on the tree lattice can be given by $\sqrt{z}\epsilon_n$, which is further verified by diagonalizing the tight-binding Hamiltonian directly. In the case where $|t_1|<|\gamma|$, the Hamiltonian cannot be transferred to a Hermitian counterpart, and the spectrum can be obtained through numerical diagonalization.

Based on the above analytical solution, it become evident that the infinite-dimensional Bethe lattice can be separated into a sequence of 1D chains, with the chain length $L$ varying from $1$ to $G+1$. One end boundary of the 1D chains formed by $\psi_l$ or $\phi^\alpha_{l,r}$ with a length of $L$ is located at the sites in the outermost boundary (the $G$-th generation), while the other end boundary is located at the sites in the $(G+1-L)$-th generation within the bulk. The formation of 1D chains with different lengths on the Bethe lattice critically depends on the exchange invariance exhibited by the subtree branches generated at each node. This exchange invariance gives rise to the \emph{bulk-boundary duality} on the Bethe lattice, allowing internal positions within the bulk to act as boundaries as well. The bulk-boundary duality and non-reciprocal hopping thus are responsible for the appearance of inner non-Hermitian skin effect on Bethe lattice.

\begin{figure}[htbp]
  \centering
  \includegraphics[width=8.6 cm]{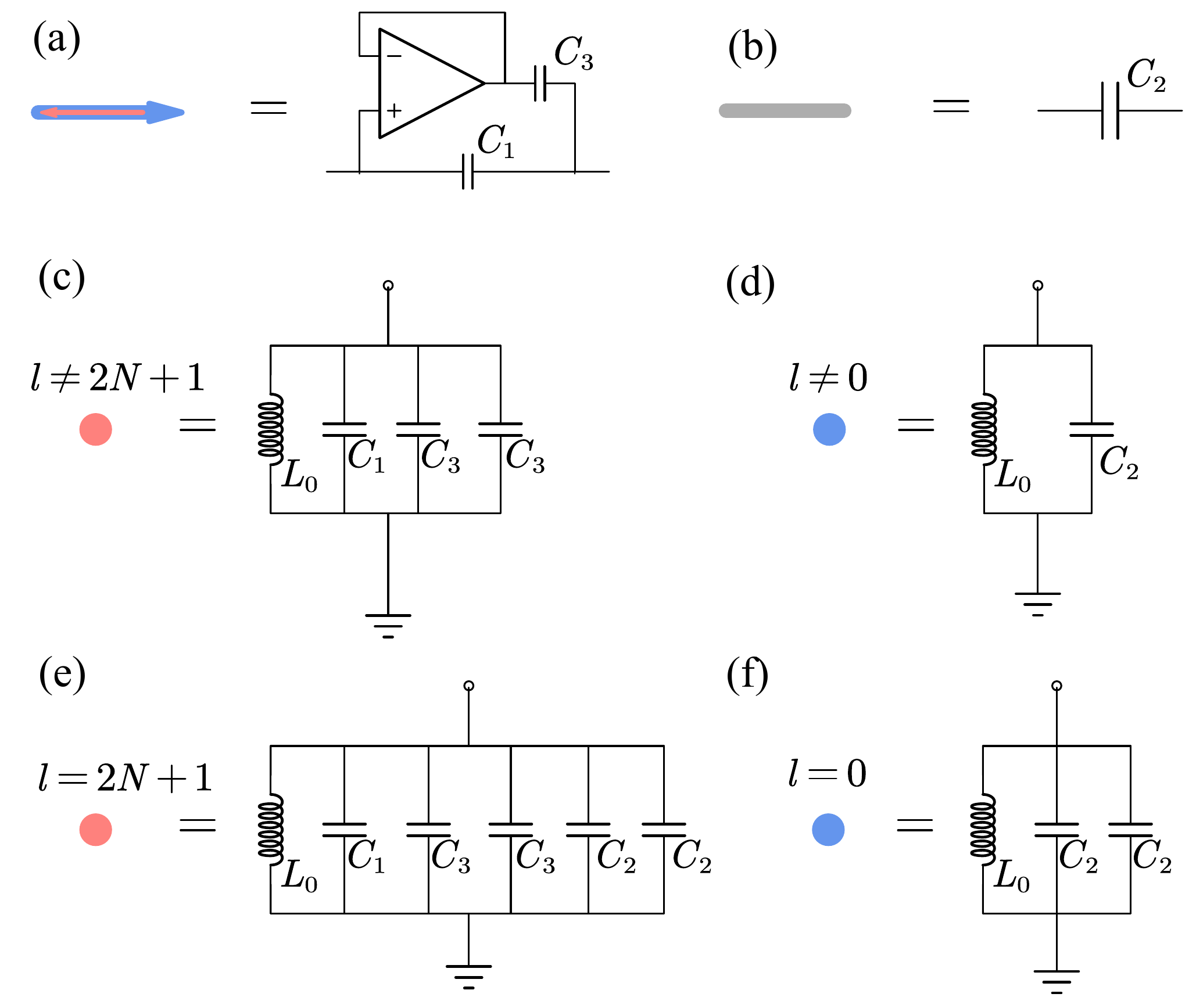}\\
  \caption{Schematic of the circuit implementations of the basic elements of the tree lattice shown in Fig.~\ref{fig1}: (a) non-reciprocal bond with the hopping amplitudes $t_1\pm \gamma$, (b) reciprocal bond with $t_2$, (c) site in the odd generation, (d) site in the even generation, (e) site in the outermost generation, and (f) the root site. }\label{fig4}
\end{figure}

\textit{Simulations of inner skin effect on electric circuit.-}
To further validate our proposal, we implement the non-Hermitian SSH model using a Bethe-lattice circuit composed of capacitors, inductors, and operational amplifiers. The non-Hermitian effects resulting from non-reciprocal hopping can be incorporated by utilizing operational amplifiers\cite{Lee2018,PhysRevResearch.3.023056,PhysRevB.103.125411,PhysRevB.103.035420,Liu2022,PhysRevResearch.5.013107}. By applying Kirchhoff's current law, we can design a circuit that corresponds to our non-Hermitian tight-binding model in Eq.~(\ref{eq1}), with the basic circuit elements depicted in Fig.~\ref{fig4}.
The Laplacian operator $\mathcal{L}$ for this circuit, which relates the current and voltage distributions in the circuit ($I = \mathcal{L}V$), can be written as:
\begin{equation}\label{}
\mathcal{L}={\rm i}\omega C_2\left\{\mathcal{H}-\left[2(t_1+\gamma)+2t_2-\frac{\omega_c^2}{\omega^2}\right]\right\}
\end{equation}
where $t_1=C_1/C_2+C_3/(2C_2)$, $t_2=C_2/C_2=1$, $\gamma=C_3/(2C_2)$ and $\omega_c=1/\sqrt{C_2L_0}$.
In the circuit simulation, we set $C_1=7\ nF$, $C_2=1\ uF$, $C_3=1.4\ uF$ and $L_0=1\ mH$. If no external power source is connected to the circuit (achieved experimentally by inputting a pulse signal to the circuit and then removing the power source), the current vector $I=\mathcal{L}V=0$. Thus, we have
\begin{equation}\label{}
\left\{\mathcal{H}-\left[2(t_1+\gamma)+2\right]\right\}V=-\frac{\omega_c^2}{\omega^2}V.
\end{equation}
It implies there is a correspondence between the resonance frequency $\omega_n$ of this circuit and the eigenvalues $E_n$ of the simulated Hamiltonian $\mathcal{H}$
\begin{equation}\label{eq_omega}
  \omega_n=\omega_c\sqrt{-\frac{1}{E_n-\left[2(t_1+\gamma)+2\right]}}.
\end{equation}
Additionally, the voltage distribution $V= \{V_{n,i}\}$ (`$n$' denotes the resonance frequency and `$i$' denotes the site) is proportional to the eigenstate of $\mathcal{H}$ with the eigenvalue $E_n$.

\begin{figure}[hbpt]
  \centering
  \includegraphics[width=8.6 cm]{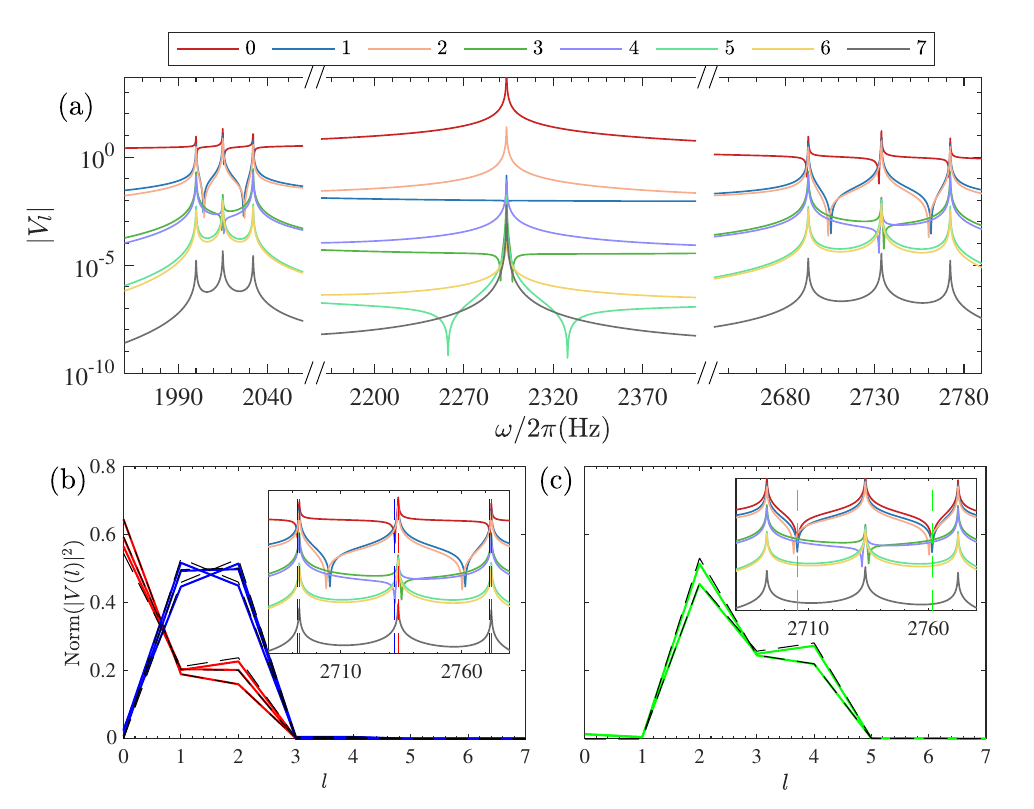}\\
  \caption{Results obtained from circuit simulations of the non-Hermitian SSH model on the Bethe lattice: (a) voltage at each node as a function of frequency due to an AC signal source located at the root node with an amplitude of $V_0 = 1 \text{ V}$. The peak and valley frequencies are consistent with the theoretical resonance frequencies of the circuit's Laplacian $\mathcal{L}$. (b) and (c) display the squared distribution of the normalized voltage at the resonance frequencies indicated in the insets. Here the black dashed lines are results obtained from direct diagonalization of the non-Hermitian SSH Hamiltonian. The simulated tree lattice with circuits consists of $G=7$ generations.}\label{fig5}
\end{figure}

We use LTspice software to simulate the circuit designed for the non-Hermitian SSH model on a Bethe lattice with $G=7$ generations. Initially, we introduce a voltage source signal at the root node and perform AC analysis of small signals. This analysis provides the frequency response of the voltages at various nodes of the circuit relative to the AC signal source. As shown in Fig.~\ref{fig5}(a), the results reveal significant voltage response with sharp peaks and valleys at specific frequencies for each node. These frequencies correspond to the circuit's resonance frequencies, consistent with those predicted by the theoretical calculations [see Eq. (\ref{eq_omega})]. 

The voltage distribution at the resonance frequency allows us to obtain information about the wavefunction of the eigenstate as well. We extract and normalize the voltage distributions at several resonance frequencies, which are shown in Figs.~\ref{fig5}~(b) and \ref{fig5}(c). The eigenstates shown in Fig.~\ref{fig5}(b) correspond to the bulk states of 1D effective chains with sizes $L=8$ (red) and $L=7$ (blue), while those in Fig.~\ref{fig5}(c) are for size $L=6$. These eigenstates exhibit the inner skin effect, localizing near the inner boundaries of the 1D effective chains. The voltage distributions are also compared to the wavefunctions obtained by directly diagonalizing the non-Hermitian SSH Hamiltonian, showing a nice consistency between them.
Therefore, we expect that the inner non-Hermitian skin effect on a Bethe lattice can be experimentally implemented and observed in circuits.

\textit{Conclusion.-}
We investigate the properties of non-Hermitian SSH model on Bethe lattice and reveal a new form of skin effect, termed the inner non-Hermitian skin effect, where all eigenstates become localized within the bulk region. The Hamiltonian on Bethe lattice permits an analytical solution and can be decoupled into a sequence of 1D chains, ranging from 1 to the size of the tree lattice. Since all chains are arranged with one end positioned at the bottom generation, the other end boundaries are naturally located at varying generations within the bulk, depending on the lengths of the chains. It is precisely these inner ends that result in the inner skin effect. Finally, we conduct circuit simulations and predict that this unusual skin effect can be observed in artificial experimental platforms, such as circuits.

The authors are indebted to L. Li, W. Zhang, X. Zhang and X.-X. Zhang for insightful discussions. J.S and H.G. acknowledges support from the NSFC grant No.~12074022. S.F. is supported by the National Key Research and Development Program of China under Grant Nos. 2023YFA1406500 and 2021YFA1401803, and NSFC under Grant No. 12274036.

\textit{Note added.-} After completing the present manuscript, we came to notice related works on arXiv:2408.11024\cite{multifractal} and arXiv:2409.01873\cite{quantumtransportbethe}.

\appendix
\setcounter{equation}{0}
\renewcommand\theequation{A.\arabic{equation}}
\setcounter{figure}{0}
\renewcommand{\thefigure}{A\arabic{figure}}

\section{Details of the analytical solution}\label{appB}
\begin{figure}[h]
  \centering
  \includegraphics[width=6. cm]{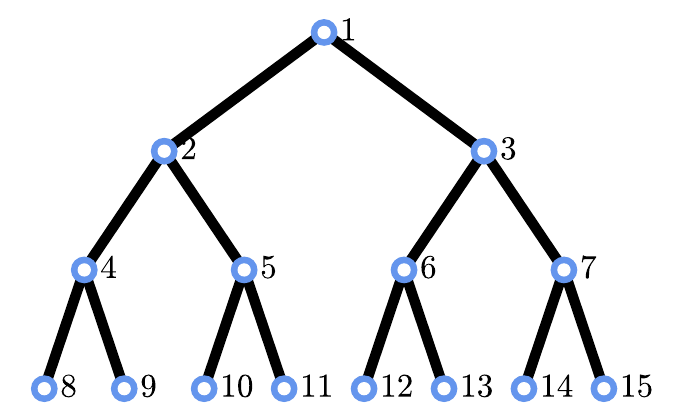}\\
  \caption{Schematic demonstration of a tree lattice with $G=3$ generations.}\label{afig2}
\end{figure}
In this appendix, we use the tree lattice with $G=3$ generations (consisting of $15$ sites) as an example to provide the details of the analytical solution. 

Under the basis $X=\{c_1,c_2,\cdots,c_{15}\}^T$, the tight-binding Hamiltonian of the non-Hermitian SSH model on the tree lattice can be expressed as $H=X^{\dagger}\mathcal{H}X$, where $\mathcal{H}$ is the Hamiltonian matrix. Assuming the eigenfunction has the form $|\Psi\rangle=(\phi_1,\phi_2,\cdots,\phi_{15})^T$, the eigenproblem $\mathcal{H}|\Psi\rangle=E|\Psi\rangle$ can be expanded as follows:
\begin{equation}\label{eqcomponent}
\begin{split}
&E\phi_1=(t_1+\gamma)(\phi_2+\phi_3),\\
&E\phi_2=(t_1-\gamma)\phi_1+t_2(\phi_4+\phi_5),\\
&E\phi_3=(t_1-\gamma)\phi_1+t_2(\phi_6+\phi_7),\\
&E\phi_4=t_2\phi_2+(t_1+\gamma)(\phi_8+\phi_9),\\
&E\phi_5=t_2\phi_2+(t_1+\gamma)(\phi_{10}+\phi_{11}),\\
&E\phi_6=t_2\phi_3+(t_1+\gamma)(\phi_{12}+\phi_{13}),\\
&E\phi_7=t_2\phi_3+(t_1+\gamma)(\phi_{14}+\phi_{15}),\\
&E\phi_8=(t_1-\gamma)\phi_4,E\phi_9=(t_1-\gamma)\phi_4,\\
&E\phi_{10}=(t_1-\gamma)\phi_5,E\phi_{11}=(t_1-\gamma)\phi_5,\\
&E\phi_{12}=(t_1-\gamma)\phi_6,E\phi_{13}=(t_1-\gamma)\phi_6,\\
&E\phi_{14}=(t_1-\gamma)\phi_7,E\phi_{15}=(t_1-\gamma)\phi_7.
\end{split}
\end{equation}
We define a set of linear combinations of the basis wavefunctions as follows
\begin{equation}\label{}
\begin{split}
\psi_l=\frac{1}{\sqrt{z^l}}\sum_{i=1}^{N_l}\phi_i,
\end{split}
\end{equation}
where the summation is taken over the $N_l=z^{l}$ $(z=2)$ sites in the $l$-th generation. Based on the site labels in Fig.~\ref{afig2}(a), $\psi_{l}$ can be explicitly written as:
\begin{equation}\label{}
\begin{split}
&\psi_{0}=\phi_1,\psi_1=\frac1{\sqrt2}(\phi_2+\phi_3),\\
&\psi_2=\frac1{\sqrt {2^2}}\sum_{i=4}^{7}\phi_i,\psi_3=\frac1{\sqrt {2^3}}\sum_{i=8}^{15}\phi_i.
\end{split}
\end{equation}
In terms of $\psi_l$, we can obtain the following four equations by summing the equations in Eq.(\ref{eqcomponent}) that include $\phi_i$ on the left-hand side belonging to the same generation:
\begin{equation}\label{lenG}
\begin{split}
&E\psi_0=\sqrt{z}(t_1+\gamma)\psi_1,\\
&E\psi_1=\sqrt{z}\{(t_1-\gamma)\psi_0+t_2\psi_2\},\\
&E\psi_2=\sqrt{z}\{t_2\psi_1+(t_1+\gamma)\psi_3\},\\
&E\psi_4=\sqrt{z}(t_1-\gamma)\psi_3.
\end{split}
\end{equation}

Another set of linear combinations of wavesfunctions containing sites on different branches in the $r$-th($r=1,\cdots,G-l$) generation, generated from the $\alpha$ site in the $l$-th($l=0,1,\cdots,G$) generation, are defined as
\begin{equation}\label{}
\begin{split}
\phi^{l,\alpha}_r=\frac{1}{\sqrt{N^{i,\alpha}_{r}(\sum_{i=1}^z|C_i|^2})}\sum_{i=1}^zC_i\left(\sum_{j=1}^{N^{i,\alpha}_{r}}\phi_j\right),
\end{split}
\end{equation}
where the first summation is made over the $z$ sub-branches starting from the reference site $\alpha$; the second summation is taken over the $N^{i,\alpha}_{r}$ sites in the $r$-th generation along the $i$-th sub-branch starting from the reference site $\alpha$ in the $l$-th generation; $C_i$ represents the coefficient of the $i$-th sub-branch. $C_i$ satisfies the condition of $\sum_i^zC_i=0$, which is essential as it restricts the presence of only $z-1$ sets of linearly independent solutions. Consequently, the degeneracy of $\phi_{r}^{l,\alpha}$ for every $(l,r)$ pair is $z-1$.
For the finite lattice in Fig.\ref{afig2}(a), $\phi^{l,\alpha}_r$ can be explicitly expressed as:
\begin{equation}\label{}
\begin{split}
&\phi^{0,1}_1=\frac{1}{\sqrt2}(\phi_2-\phi_3),\\
&\phi^{0,1}_2=\frac{1}{\sqrt4}\left[(\phi_4+\phi_5)-(\phi_6+\phi_7)\right],\\
&\phi^{0,1}_3=\frac{1}{\sqrt8}\left[\left(\sum_{i=8}^{11}\phi_i\right)-\left(\sum_{i=12}^{15}\phi_i\right)\right],\\
&\phi^{1,2}_1=\frac{1}{\sqrt2}(\phi_4-\phi_5),\phi^{1,3}_1=\frac{1}{\sqrt2}(\phi_6-\phi_7),\\
&\phi^{1,2}_2=\frac{1}{\sqrt4}\left[(\phi_8+\phi_9)-(\phi_{10}+\phi_{11})\right],\\
&\phi^{1,3}_2=\frac{1}{\sqrt4}\left[(\phi_{12}+\phi_{13})-(\phi_{14}+\phi_{15})\right],\\
&\phi^{2,4}_1=\frac{1}{\sqrt2}(\phi_8-\phi_9),\phi^{2,5}_1=\frac{1}{\sqrt2}(\phi_{10}-\phi_{11}),\\
&\phi^{2,6}_1=\frac{1}{\sqrt2}(\phi_{12}-\phi_{13}),\phi^{2,7}_1=\frac{1}{\sqrt2}(\phi_{14}-\phi_{15}).\\
\end{split}
\end{equation}

In terms of $\phi^{l,\alpha}_{r}$, we can derive the following eleven equations from Eq.(\ref{eqcomponent}):
\begin{equation}\label{lenL}
\begin{split}
&E\phi_1^{0,1}=\sqrt{z}t_2\phi_2^{0,1},\\
&E\phi_2^{0,1}=\sqrt{z}\{t_2\phi_1^{0,1}+(t_1+\gamma)\phi_3^{0,1}\},\\
&E\phi_3^{0,1}=\sqrt{z}(t_1-\gamma)\phi_2^{0,1},\\
&E\phi_1^{1,2}=\sqrt{z}(t_1+\gamma)\phi_2^{1,2},\\
&E\phi_2^{1,2}=\sqrt{z}(t_1-\gamma)\phi_1^{1,2},\\
&E\phi_1^{1,3}=\sqrt{z}(t_1+\gamma)\phi_2^{1,3},\\
&E\phi_2^{1,3}=\sqrt{z}(t_1-\gamma)\phi_1^{1,3},\\
&E\phi_1^{2,4}=0,E\phi_1^{2,5}=0,\\
&E\phi_1^{2,6}=0,E\phi_1^{2,7}=0.
\end{split}
\end{equation}

\begin{figure}[t]
  \centering
  \includegraphics[width=8 cm]{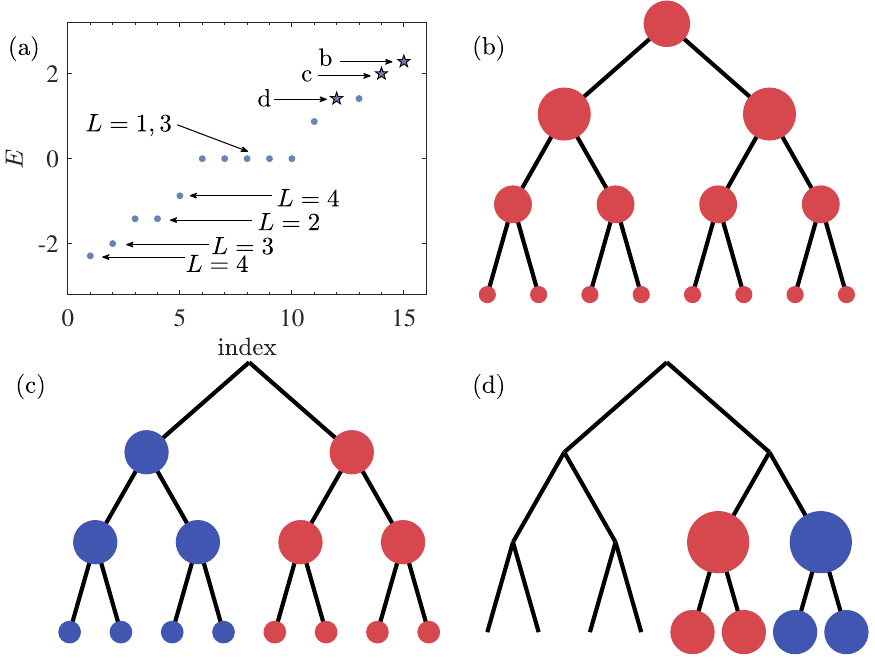}\\
  \caption{The eigenvalues and eigenvectors of the tight-binding model on a tree lattice with $G=3$ generations. The hopping amplitudes are set as $t_1=t_2=1$ and $\gamma=0$ for demonstration purpose. (a): Eigenvalues, each of which is associated with its corresponding chain length. (b), (c), and (d): Wavefunctions of the eigenstates labeled as b, c, and d in (a), respectively. In (b), (c), and (d), the red (blue) color represents positive (negative) amplitudes, with their absolute values indicated by the marker size.}\label{afig3}
\end{figure}

It is evident that Eqs. (\ref{lenG}) and (\ref{lenL}), expressed in terms of $\{\psi_l\}$ and $\{\phi_r^{l,\alpha}\}$, correspond to the eigenproblems of a set of 1D non-Hermitian SSH chains with the alternating hopping amplitudes of $\sqrt{z}(t_1\pm\gamma)$ and $\sqrt{z}t_2$. The equations in Eqs. (\ref{lenG}) correspond to a chain with $L=4$ sites, where the basis is given by $\{\psi_l\}$. The equations in Eq.(\ref{lenL}) can be categorized according to the values of the $(l,\alpha)$ pair. Each set of equations sharing the same $(l,\alpha)$ values represents a 1D non-Hermitian SSH chain, with its length determined by the number of equations in that particular set. For example, the chain length of $\{\phi^{0,1}_{r}\}$ is $L=3$. The chains $\{\phi^{1,2}_{r}\}$ and $\{\phi^{1,3}_{r}\}$ have a length of $L=2$, representing two degenerate chains. On the other hand, the chains $\{\phi^{2,4}_{1}\}, \{\phi^{2,5}_{1}\}, \{\phi^{2,6}_{1}\}$, and $\{\phi^{2,7}_{1}\}$ constitute four degenerate chains with a length of $L=1$. Generally, for a non-Hermitian SSH model on a tree lattice with $z$ branches and $G$ generations, the set $\{\psi_l\}$ forms a chain of length $G+1$, while the set $\{\phi^{l,\alpha}_r\}$ consists of $G$ distinct chains with lengths varying from $1$ to $G$. The degeneracy for each length is given by $N_{D}(L)=(z-1)z^{G-L}$ for $L\leq G$.

To verify the analytical solution, we perform direct diagonalization of the tight-binding Hamiltonian on a tree lattice. The resulting number and values of the eigenvalues are exactly the same with those obtained from decoupled 1D chains [see Fig.\ref{afig3}(a)]. The chain to which each eigenvalue belongs can then be identified, and the distribution of the corresponding eigenstate is consistent with the expectation.
Figure \ref{afig3}(b) depicts the eigenstate associated with a chain length of $L=4$. The eigenstate is distributed across all generations, and the amplitude is symmetric across the sites within each generation. In Fig. \ref{afig3}(c), the wavefunction of an eigenstate for the $L=3$ chain is depicted. Indeed, this particular eigenstate is found to be only distributed throughout the outermost $L=3$ generations, with its amplitudes satisfying the following constraint: values within the same generation are equal, but their sum equals zero. As illustrated in Fig. \ref{afig3}(d), the wavefunction of the eigenstate corresponding to an $L=2$ chain exhibits a similar pattern. This eigenstate is only spread across the outermost $L=2$ generations, where amplitudes within the same generation are equal and sum up to zero.

\bibliographystyle{apsrev4-1-etal-title_10authors}
\bibliography{bethe_ref}

\begin{thebibliography}{65}%
\makeatletter
\providecommand \@ifxundefined [1]{%
 \@ifx{#1\undefined}
}%
\providecommand \@ifnum [1]{%
 \ifnum #1\expandafter \@firstoftwo
 \else \expandafter \@secondoftwo
 \fi
}%
\providecommand \@ifx [1]{%
 \ifx #1\expandafter \@firstoftwo
 \else \expandafter \@secondoftwo
 \fi
}%
\providecommand \natexlab [1]{#1}%
\providecommand \enquote  [1]{``#1''}%
\providecommand \bibnamefont  [1]{#1}%
\providecommand \bibfnamefont [1]{#1}%
\providecommand \citenamefont [1]{#1}%
\providecommand \href@noop [0]{\@secondoftwo}%
\providecommand \href [0]{\begingroup \@sanitize@url \@href}%
\providecommand \@href[1]{\@@startlink{#1}\@@href}%
\providecommand \@@href[1]{\endgroup#1\@@endlink}%
\providecommand \@sanitize@url [0]{\catcode `\\12\catcode `\$12\catcode `\&12\catcode `\#12\catcode `\^12\catcode `\_12\catcode `\%12\relax}%
\providecommand \@@startlink[1]{}%
\providecommand \@@endlink[0]{}%
\providecommand \url  [0]{\begingroup\@sanitize@url \@url }%
\providecommand \@url [1]{\endgroup\@href {#1}{\urlprefix }}%
\providecommand \urlprefix  [0]{URL }%
\providecommand \Eprint [0]{\href }%
\providecommand \doibase [0]{http://dx.doi.org/}%
\providecommand \selectlanguage [0]{\@gobble}%
\providecommand \bibinfo  [0]{\@secondoftwo}%
\providecommand \bibfield  [0]{\@secondoftwo}%
\providecommand \translation [1]{[#1]}%
\providecommand \BibitemOpen [0]{}%
\providecommand \bibitemStop [0]{}%
\providecommand \bibitemNoStop [0]{.\EOS\space}%
\providecommand \EOS [0]{\spacefactor3000\relax}%
\providecommand \BibitemShut  [1]{\csname bibitem#1\endcsname}%
\let\auto@bib@innerbib\@empty
\bibitem [{\citenamefont {Mahan}(2001)}]{PhysRevB.63.155110}%
  \BibitemOpen
  \bibfield  {author} {\bibinfo {author} {\bibfnamefont {G.~D.}\ \bibnamefont {Mahan}},\ }\bibfield  {title} {\enquote {\bibinfo {title} {Energy bands of the {Bethe} lattice}}, }\href {\doibase 10.1103/PhysRevB.63.155110} {\bibfield  {journal} {\bibinfo  {journal} {Phys. Rev. B}\ }\textbf {\bibinfo {volume} {63}},\ \bibinfo {pages} {155110} (\bibinfo {year} {2001})}\BibitemShut {NoStop}%
\bibitem [{\citenamefont {Aryal}\ and\ \citenamefont {Kettemann}(2020)}]{Aryal_2020}%
  \BibitemOpen
  \bibfield  {author} {\bibinfo {author} {\bibfnamefont {D.}~\bibnamefont {Aryal}}\ and\ \bibinfo {author} {\bibfnamefont {S.}~\bibnamefont {Kettemann}},\ }\bibfield  {title} {\enquote {\bibinfo {title} {Complete solution of the tight binding model on a {Cayley} tree: strongly localised versus extended states}}, }\href {\doibase 10.1088/2399-6528/abc1c3} {\bibfield  {journal} {\bibinfo  {journal} {Journal of Physics Communications}\ }\textbf {\bibinfo {volume} {4}},\ \bibinfo {pages} {105010} (\bibinfo {year} {2020})}\BibitemShut {NoStop}%
\bibitem [{\citenamefont {da~Silva}\ and\ \citenamefont {Coutinho}(1986)}]{PhysRevB.34.7975}%
  \BibitemOpen
  \bibfield  {author} {\bibinfo {author} {\bibfnamefont {C.~R.}\ \bibnamefont {da~Silva}}\ and\ \bibinfo {author} {\bibfnamefont {S.}~\bibnamefont {Coutinho}},\ }\bibfield  {title} {\enquote {\bibinfo {title} {{Ising} model on the {Bethe} lattice with competing interactions up to the third-nearest-neighbor generation}}, }\href {\doibase 10.1103/PhysRevB.34.7975} {\bibfield  {journal} {\bibinfo  {journal} {Phys. Rev. B}\ }\textbf {\bibinfo {volume} {34}},\ \bibinfo {pages} {7975} (\bibinfo {year} {1986})}\BibitemShut {NoStop}%
\bibitem [{\citenamefont {Gujrati}(1995)}]{PhysRevLett.74.809}%
  \BibitemOpen
  \bibfield  {author} {\bibinfo {author} {\bibfnamefont {P.~D.}\ \bibnamefont {Gujrati}},\ }\bibfield  {title} {\enquote {\bibinfo {title} {Bethe or {Bethe}-like {Lattice Calculations Are More Reliable Than Conventional Mean-Field Calculations}}}, }\href {\doibase 10.1103/PhysRevLett.74.809} {\bibfield  {journal} {\bibinfo  {journal} {Phys. Rev. Lett.}\ }\textbf {\bibinfo {volume} {74}},\ \bibinfo {pages} {809} (\bibinfo {year} {1995})}\BibitemShut {NoStop}%
\bibitem [{\citenamefont {Eckstein}\ \emph {et~al.}(2005)\citenamefont {Eckstein}, \citenamefont {Kollar}, \citenamefont {Byczuk},\ and\ \citenamefont {Vollhardt}}]{PhysRevB.71.235119}%
  \BibitemOpen
  \bibfield  {author} {\bibinfo {author} {\bibfnamefont {M.}~\bibnamefont {Eckstein}}, \bibinfo {author} {\bibfnamefont {M.}~\bibnamefont {Kollar}}, \bibinfo {author} {\bibfnamefont {K.}~\bibnamefont {Byczuk}}, \ and\ \bibinfo {author} {\bibfnamefont {D.}~\bibnamefont {Vollhardt}},\ }\bibfield  {title} {\enquote {\bibinfo {title} {Hopping on the {Bethe} lattice: {Exact} results for densities of states and dynamical mean-field theory}}, }\href {\doibase 10.1103/PhysRevB.71.235119} {\bibfield  {journal} {\bibinfo  {journal} {Phys. Rev. B}\ }\textbf {\bibinfo {volume} {71}},\ \bibinfo {pages} {235119} (\bibinfo {year} {2005})}\BibitemShut {NoStop}%
\bibitem [{\citenamefont {Thouless}(1986)}]{PhysRevLett.56.1082}%
  \BibitemOpen
  \bibfield  {author} {\bibinfo {author} {\bibfnamefont {D.~J.}\ \bibnamefont {Thouless}},\ }\bibfield  {title} {\enquote {\bibinfo {title} {Spin-{Glass} on a {Bethe Lattice}}}, }\href {\doibase 10.1103/PhysRevLett.56.1082} {\bibfield  {journal} {\bibinfo  {journal} {Phys. Rev. Lett.}\ }\textbf {\bibinfo {volume} {56}},\ \bibinfo {pages} {1082} (\bibinfo {year} {1986})}\BibitemShut {NoStop}%
\bibitem [{\citenamefont {Tragtenberg}\ and\ \citenamefont {Yokoi}(1995)}]{PhysRevE.52.2187}%
  \BibitemOpen
  \bibfield  {author} {\bibinfo {author} {\bibfnamefont {M.~H.~R.}\ \bibnamefont {Tragtenberg}}\ and\ \bibinfo {author} {\bibfnamefont {C.~S.~O.}\ \bibnamefont {Yokoi}},\ }\bibfield  {title} {\enquote {\bibinfo {title} {Field behavior of an {Ising} model with competing interactions on the {Bethe} lattice}}, }\href {\doibase 10.1103/PhysRevE.52.2187} {\bibfield  {journal} {\bibinfo  {journal} {Phys. Rev. E}\ }\textbf {\bibinfo {volume} {52}},\ \bibinfo {pages} {2187} (\bibinfo {year} {1995})}\BibitemShut {NoStop}%
\bibitem [{\citenamefont {Laumann}\ \emph {et~al.}(2008)\citenamefont {Laumann}, \citenamefont {Scardicchio},\ and\ \citenamefont {Sondhi}}]{PhysRevB.78.134424}%
  \BibitemOpen
  \bibfield  {author} {\bibinfo {author} {\bibfnamefont {C.}~\bibnamefont {Laumann}}, \bibinfo {author} {\bibfnamefont {A.}~\bibnamefont {Scardicchio}}, \ and\ \bibinfo {author} {\bibfnamefont {S.~L.}\ \bibnamefont {Sondhi}},\ }\bibfield  {title} {\enquote {\bibinfo {title} {Cavity method for quantum spin glasses on the {Bethe} lattice}}, }\href {\doibase 10.1103/PhysRevB.78.134424} {\bibfield  {journal} {\bibinfo  {journal} {Phys. Rev. B}\ }\textbf {\bibinfo {volume} {78}},\ \bibinfo {pages} {134424} (\bibinfo {year} {2008})}\BibitemShut {NoStop}%
\bibitem [{\citenamefont {Semerjian}\ \emph {et~al.}(2009)\citenamefont {Semerjian}, \citenamefont {Tarzia},\ and\ \citenamefont {Zamponi}}]{PhysRevB.80.014524}%
  \BibitemOpen
  \bibfield  {author} {\bibinfo {author} {\bibfnamefont {G.}~\bibnamefont {Semerjian}}, \bibinfo {author} {\bibfnamefont {M.}~\bibnamefont {Tarzia}}, \ and\ \bibinfo {author} {\bibfnamefont {F.}~\bibnamefont {Zamponi}},\ }\bibfield  {title} {\enquote {\bibinfo {title} {Exact solution of the {Bose}-{Hubbard} model on the bethe lattice}}, }\href {\doibase 10.1103/PhysRevB.80.014524} {\bibfield  {journal} {\bibinfo  {journal} {Phys. Rev. B}\ }\textbf {\bibinfo {volume} {80}},\ \bibinfo {pages} {014524} (\bibinfo {year} {2009})}\BibitemShut {NoStop}%
\bibitem [{\citenamefont {Laumann}\ \emph {et~al.}(2009)\citenamefont {Laumann}, \citenamefont {Parameswaran},\ and\ \citenamefont {Sondhi}}]{PhysRevB.80.144415}%
  \BibitemOpen
  \bibfield  {author} {\bibinfo {author} {\bibfnamefont {C.~R.}\ \bibnamefont {Laumann}}, \bibinfo {author} {\bibfnamefont {S.~A.}\ \bibnamefont {Parameswaran}}, \ and\ \bibinfo {author} {\bibfnamefont {S.~L.}\ \bibnamefont {Sondhi}},\ }\bibfield  {title} {\enquote {\bibinfo {title} {Absence of {Goldstone} bosons on the {Bethe} lattice}}, }\href {\doibase 10.1103/PhysRevB.80.144415} {\bibfield  {journal} {\bibinfo  {journal} {Phys. Rev. B}\ }\textbf {\bibinfo {volume} {80}},\ \bibinfo {pages} {144415} (\bibinfo {year} {2009})}\BibitemShut {NoStop}%
\bibitem [{\citenamefont {Li}\ \emph {et~al.}(2012)\citenamefont {Li}, \citenamefont {von Delft},\ and\ \citenamefont {Xiang}}]{PhysRevB.86.195137}%
  \BibitemOpen
  \bibfield  {author} {\bibinfo {author} {\bibfnamefont {W.}~\bibnamefont {Li}}, \bibinfo {author} {\bibfnamefont {J.}~\bibnamefont {von Delft}}, \ and\ \bibinfo {author} {\bibfnamefont {T.}~\bibnamefont {Xiang}},\ }\bibfield  {title} {\enquote {\bibinfo {title} {Efficient simulation of infinite tree tensor network states on the {Bethe} lattice}}, }\href {\doibase 10.1103/PhysRevB.86.195137} {\bibfield  {journal} {\bibinfo  {journal} {Phys. Rev. B}\ }\textbf {\bibinfo {volume} {86}},\ \bibinfo {pages} {195137} (\bibinfo {year} {2012})}\BibitemShut {NoStop}%
\bibitem [{\citenamefont {Daniška}\ and\ \citenamefont {Gendiar}(2016)}]{Daniska2016}%
  \BibitemOpen
  \bibfield  {author} {\bibinfo {author} {\bibfnamefont {M.}~\bibnamefont {Daniška}}\ and\ \bibinfo {author} {\bibfnamefont {A.}~\bibnamefont {Gendiar}},\ }\bibfield  {title} {\enquote {\bibinfo {title} {Analysis of quantum spin models on hyperbolic lattices and {Bethe} lattice}}, }\href {\doibase 10.1088/1751-8113/49/14/145003} {\bibfield  {journal} {\bibinfo  {journal} {Journal of Physics A: Mathematical and Theoretical}\ }\textbf {\bibinfo {volume} {49}},\ \bibinfo {pages} {145003} (\bibinfo {year} {2016})}\BibitemShut {NoStop}%
\bibitem [{\citenamefont {Qu}\ \emph {et~al.}(2019)\citenamefont {Qu}, \citenamefont {Li},\ and\ \citenamefont {Xiang}}]{PhysRevB.100.125121}%
  \BibitemOpen
  \bibfield  {author} {\bibinfo {author} {\bibfnamefont {D.-W.}\ \bibnamefont {Qu}}, \bibinfo {author} {\bibfnamefont {W.}~\bibnamefont {Li}}, \ and\ \bibinfo {author} {\bibfnamefont {T.}~\bibnamefont {Xiang}},\ }\bibfield  {title} {\enquote {\bibinfo {title} {Thermal tensor network simulations of the {Heisenberg} model on the {Bethe} lattice}}, }\href {\doibase 10.1103/PhysRevB.100.125121} {\bibfield  {journal} {\bibinfo  {journal} {Phys. Rev. B}\ }\textbf {\bibinfo {volume} {100}},\ \bibinfo {pages} {125121} (\bibinfo {year} {2019})}\BibitemShut {NoStop}%
\bibitem [{\citenamefont {Lunts}\ \emph {et~al.}(2021)\citenamefont {Lunts}, \citenamefont {Georges}, \citenamefont {Stoudenmire},\ and\ \citenamefont {Fishman}}]{PhysRevResearch.3.023054}%
  \BibitemOpen
  \bibfield  {author} {\bibinfo {author} {\bibfnamefont {P.}~\bibnamefont {Lunts}}, \bibinfo {author} {\bibfnamefont {A.}~\bibnamefont {Georges}}, \bibinfo {author} {\bibfnamefont {E.~M.}\ \bibnamefont {Stoudenmire}}, \ and\ \bibinfo {author} {\bibfnamefont {M.}~\bibnamefont {Fishman}},\ }\bibfield  {title} {\enquote {\bibinfo {title} {Hubbard model on the {Bethe} lattice via variational uniform tree states: {Metal}-insulator transition and a {Fermi} liquid}}, }\href {\doibase 10.1103/PhysRevResearch.3.023054} {\bibfield  {journal} {\bibinfo  {journal} {Phys. Rev. Res.}\ }\textbf {\bibinfo {volume} {3}},\ \bibinfo {pages} {023054} (\bibinfo {year} {2021})}\BibitemShut {NoStop}%
\bibitem [{\citenamefont {Zirnbauer}(1986)}]{PhysRevB.34.6394}%
  \BibitemOpen
  \bibfield  {author} {\bibinfo {author} {\bibfnamefont {M.~R.}\ \bibnamefont {Zirnbauer}},\ }\bibfield  {title} {\enquote {\bibinfo {title} {Localization transition on the {Bethe} lattice}}, }\href {\doibase 10.1103/PhysRevB.34.6394} {\bibfield  {journal} {\bibinfo  {journal} {Phys. Rev. B}\ }\textbf {\bibinfo {volume} {34}},\ \bibinfo {pages} {6394} (\bibinfo {year} {1986})}\BibitemShut {NoStop}%
\bibitem [{\citenamefont {Mirlin}\ and\ \citenamefont {Fyodorov}(1994)}]{PhysRevLett.72.526}%
  \BibitemOpen
  \bibfield  {author} {\bibinfo {author} {\bibfnamefont {A.~D.}\ \bibnamefont {Mirlin}}\ and\ \bibinfo {author} {\bibfnamefont {Y.~V.}\ \bibnamefont {Fyodorov}},\ }\bibfield  {title} {\enquote {\bibinfo {title} {Distribution of local densities of states, order parameter function, and critical behavior near the {Anderson} transition}}, }\href {\doibase 10.1103/PhysRevLett.72.526} {\bibfield  {journal} {\bibinfo  {journal} {Phys. Rev. Lett.}\ }\textbf {\bibinfo {volume} {72}},\ \bibinfo {pages} {526} (\bibinfo {year} {1994})}\BibitemShut {NoStop}%
\bibitem [{\citenamefont {De~Luca}\ \emph {et~al.}(2014)\citenamefont {De~Luca}, \citenamefont {Altshuler}, \citenamefont {Kravtsov},\ and\ \citenamefont {Scardicchio}}]{PhysRevLett.113.046806}%
  \BibitemOpen
  \bibfield  {author} {\bibinfo {author} {\bibfnamefont {A.}~\bibnamefont {De~Luca}}, \bibinfo {author} {\bibfnamefont {B.~L.}\ \bibnamefont {Altshuler}}, \bibinfo {author} {\bibfnamefont {V.~E.}\ \bibnamefont {Kravtsov}}, \ and\ \bibinfo {author} {\bibfnamefont {A.}~\bibnamefont {Scardicchio}},\ }\bibfield  {title} {\enquote {\bibinfo {title} {{Anderson Localization} on the {Bethe Lattice}: {Nonergodicity} of {Extended States}}}, }\href {\doibase 10.1103/PhysRevLett.113.046806} {\bibfield  {journal} {\bibinfo  {journal} {Phys. Rev. Lett.}\ }\textbf {\bibinfo {volume} {113}},\ \bibinfo {pages} {046806} (\bibinfo {year} {2014})}\BibitemShut {NoStop}%
\bibitem [{\citenamefont {Savitz}\ \emph {et~al.}(2019)\citenamefont {Savitz}, \citenamefont {Peng},\ and\ \citenamefont {Refael}}]{PhysRevB.100.094201}%
  \BibitemOpen
  \bibfield  {author} {\bibinfo {author} {\bibfnamefont {S.}~\bibnamefont {Savitz}}, \bibinfo {author} {\bibfnamefont {C.}~\bibnamefont {Peng}}, \ and\ \bibinfo {author} {\bibfnamefont {G.}~\bibnamefont {Refael}},\ }\bibfield  {title} {\enquote {\bibinfo {title} {Anderson localization on the {Bethe} lattice using cages and the {Wegner} flow}}, }\href {\doibase 10.1103/PhysRevB.100.094201} {\bibfield  {journal} {\bibinfo  {journal} {Phys. Rev. B}\ }\textbf {\bibinfo {volume} {100}},\ \bibinfo {pages} {094201} (\bibinfo {year} {2019})}\BibitemShut {NoStop}%
\bibitem [{\citenamefont {Yao}\ and\ \citenamefont {Wang}(2018)}]{PhysRevLett.121.086803}%
  \BibitemOpen
  \bibfield  {author} {\bibinfo {author} {\bibfnamefont {S.}~\bibnamefont {Yao}}\ and\ \bibinfo {author} {\bibfnamefont {Z.}~\bibnamefont {Wang}},\ }\bibfield  {title} {\enquote {\bibinfo {title} {{Edge States} and {Topological Invariants} of {Non}-{Hermitian Systems}}}, }\href {\doibase 10.1103/PhysRevLett.121.086803} {\bibfield  {journal} {\bibinfo  {journal} {Phys. Rev. Lett.}\ }\textbf {\bibinfo {volume} {121}},\ \bibinfo {pages} {086803} (\bibinfo {year} {2018})}\BibitemShut {NoStop}%
\bibitem [{\citenamefont {Lieu}(2018)}]{PhysRevB.97.045106}%
  \BibitemOpen
  \bibfield  {author} {\bibinfo {author} {\bibfnamefont {S.}~\bibnamefont {Lieu}},\ }\bibfield  {title} {\enquote {\bibinfo {title} {Topological phases in the non-hermitian su-schrieffer-heeger model}}, }\href {\doibase 10.1103/PhysRevB.97.045106} {\bibfield  {journal} {\bibinfo  {journal} {Phys. Rev. B}\ }\textbf {\bibinfo {volume} {97}},\ \bibinfo {pages} {045106} (\bibinfo {year} {2018})}\BibitemShut {NoStop}%
\bibitem [{\citenamefont {Yao}\ \emph {et~al.}(2018)\citenamefont {Yao}, \citenamefont {Song},\ and\ \citenamefont {Wang}}]{PhysRevLett.121.136802}%
  \BibitemOpen
  \bibfield  {author} {\bibinfo {author} {\bibfnamefont {S.}~\bibnamefont {Yao}}, \bibinfo {author} {\bibfnamefont {F.}~\bibnamefont {Song}}, \ and\ \bibinfo {author} {\bibfnamefont {Z.}~\bibnamefont {Wang}},\ }\bibfield  {title} {\enquote {\bibinfo {title} {Non-hermitian chern bands}}, }\href {\doibase 10.1103/PhysRevLett.121.136802} {\bibfield  {journal} {\bibinfo  {journal} {Phys. Rev. Lett.}\ }\textbf {\bibinfo {volume} {121}},\ \bibinfo {pages} {136802} (\bibinfo {year} {2018})}\BibitemShut {NoStop}%
\bibitem [{\citenamefont {Gong}\ \emph {et~al.}(2018)\citenamefont {Gong}, \citenamefont {Ashida}, \citenamefont {Kawabata}, \citenamefont {Takasan}, \citenamefont {Higashikawa},\ and\ \citenamefont {Ueda}}]{GongZ18prx}%
  \BibitemOpen
  \bibfield  {author} {\bibinfo {author} {\bibfnamefont {Z.}~\bibnamefont {Gong}}, \bibinfo {author} {\bibfnamefont {Y.}~\bibnamefont {Ashida}}, \bibinfo {author} {\bibfnamefont {K.}~\bibnamefont {Kawabata}}, \bibinfo {author} {\bibfnamefont {K.}~\bibnamefont {Takasan}}, \bibinfo {author} {\bibfnamefont {S.}~\bibnamefont {Higashikawa}}, \ and\ \bibinfo {author} {\bibfnamefont {M.}~\bibnamefont {Ueda}},\ }\bibfield  {title} {\enquote {\bibinfo {title} {Topological phases of non-hermitian systems}}, }\href {\doibase 10.1103/PhysRevX.8.031079} {\bibfield  {journal} {\bibinfo  {journal} {Phys. Rev. X}\ }\textbf {\bibinfo {volume} {8}},\ \bibinfo {pages} {031079} (\bibinfo {year} {2018})}\BibitemShut {NoStop}%
\bibitem [{\citenamefont {Song}\ \emph {et~al.}(2019)\citenamefont {Song}, \citenamefont {Yao},\ and\ \citenamefont {Wang}}]{PhysRevLett.123.246801}%
  \BibitemOpen
  \bibfield  {author} {\bibinfo {author} {\bibfnamefont {F.}~\bibnamefont {Song}}, \bibinfo {author} {\bibfnamefont {S.}~\bibnamefont {Yao}}, \ and\ \bibinfo {author} {\bibfnamefont {Z.}~\bibnamefont {Wang}},\ }\bibfield  {title} {\enquote {\bibinfo {title} {Non-hermitian topological invariants in real space}}, }\href {\doibase 10.1103/PhysRevLett.123.246801} {\bibfield  {journal} {\bibinfo  {journal} {Phys. Rev. Lett.}\ }\textbf {\bibinfo {volume} {123}},\ \bibinfo {pages} {246801} (\bibinfo {year} {2019})}\BibitemShut {NoStop}%
\bibitem [{\citenamefont {Liu}\ \emph {et~al.}(2019)\citenamefont {Liu}, \citenamefont {Zhang}, \citenamefont {Ai}, \citenamefont {Gong}, \citenamefont {Kawabata}, \citenamefont {Ueda},\ and\ \citenamefont {Nori}}]{LiuT19prl}%
  \BibitemOpen
  \bibfield  {author} {\bibinfo {author} {\bibfnamefont {T.}~\bibnamefont {Liu}}, \bibinfo {author} {\bibfnamefont {Y.-R.}\ \bibnamefont {Zhang}}, \bibinfo {author} {\bibfnamefont {Q.}~\bibnamefont {Ai}}, \bibinfo {author} {\bibfnamefont {Z.}~\bibnamefont {Gong}}, \bibinfo {author} {\bibfnamefont {K.}~\bibnamefont {Kawabata}}, \bibinfo {author} {\bibfnamefont {M.}~\bibnamefont {Ueda}}, \ and\ \bibinfo {author} {\bibfnamefont {F.}~\bibnamefont {Nori}},\ }\bibfield  {title} {\enquote {\bibinfo {title} {Second-order topological phases in non-hermitian systems}}, }\href {\doibase 10.1103/PhysRevLett.122.076801} {\bibfield  {journal} {\bibinfo  {journal} {Phys. Rev. Lett.}\ }\textbf {\bibinfo {volume} {122}},\ \bibinfo {pages} {076801} (\bibinfo {year} {2019})}\BibitemShut {NoStop}%
\bibitem [{\citenamefont {Jin}\ and\ \citenamefont {Song}(2019)}]{JinL19prb}%
  \BibitemOpen
  \bibfield  {author} {\bibinfo {author} {\bibfnamefont {L.}~\bibnamefont {Jin}}\ and\ \bibinfo {author} {\bibfnamefont {Z.}~\bibnamefont {Song}},\ }\bibfield  {title} {\enquote {\bibinfo {title} {Bulk-boundary correspondence in a non-hermitian system in one dimension with chiral inversion symmetry}}, }\href {\doibase 10.1103/PhysRevB.99.081103} {\bibfield  {journal} {\bibinfo  {journal} {Phys. Rev. B}\ }\textbf {\bibinfo {volume} {99}},\ \bibinfo {pages} {081103} (\bibinfo {year} {2019})}\BibitemShut {NoStop}%
\bibitem [{\citenamefont {Kawabata}\ \emph {et~al.}(2019)\citenamefont {Kawabata}, \citenamefont {Shiozaki}, \citenamefont {Ueda},\ and\ \citenamefont {Sato}}]{Kawabata19prx}%
  \BibitemOpen
  \bibfield  {author} {\bibinfo {author} {\bibfnamefont {K.}~\bibnamefont {Kawabata}}, \bibinfo {author} {\bibfnamefont {K.}~\bibnamefont {Shiozaki}}, \bibinfo {author} {\bibfnamefont {M.}~\bibnamefont {Ueda}}, \ and\ \bibinfo {author} {\bibfnamefont {M.}~\bibnamefont {Sato}},\ }\bibfield  {title} {\enquote {\bibinfo {title} {Symmetry and topology in non-hermitian physics}}, }\href {\doibase 10.1103/PhysRevX.9.041015} {\bibfield  {journal} {\bibinfo  {journal} {Phys. Rev. X}\ }\textbf {\bibinfo {volume} {9}},\ \bibinfo {pages} {041015} (\bibinfo {year} {2019})}\BibitemShut {NoStop}%
\bibitem [{\citenamefont {Hamazaki}\ \emph {et~al.}(2019)\citenamefont {Hamazaki}, \citenamefont {Kawabata},\ and\ \citenamefont {Ueda}}]{PhysRevLett.123.090603}%
  \BibitemOpen
  \bibfield  {author} {\bibinfo {author} {\bibfnamefont {R.}~\bibnamefont {Hamazaki}}, \bibinfo {author} {\bibfnamefont {K.}~\bibnamefont {Kawabata}}, \ and\ \bibinfo {author} {\bibfnamefont {M.}~\bibnamefont {Ueda}},\ }\bibfield  {title} {\enquote {\bibinfo {title} {Non-hermitian many-body localization}}, }\href {\doibase 10.1103/PhysRevLett.123.090603} {\bibfield  {journal} {\bibinfo  {journal} {Phys. Rev. Lett.}\ }\textbf {\bibinfo {volume} {123}},\ \bibinfo {pages} {090603} (\bibinfo {year} {2019})}\BibitemShut {NoStop}%
\bibitem [{\citenamefont {Longhi}(2019)}]{PhysRevLett.122.237601}%
  \BibitemOpen
  \bibfield  {author} {\bibinfo {author} {\bibfnamefont {S.}~\bibnamefont {Longhi}},\ }\bibfield  {title} {\enquote {\bibinfo {title} {Topological {Phase Transition} in non-{Hermitian Quasicrystals}}}, }\href {\doibase 10.1103/PhysRevLett.122.237601} {\bibfield  {journal} {\bibinfo  {journal} {Phys. Rev. Lett.}\ }\textbf {\bibinfo {volume} {122}},\ \bibinfo {pages} {237601} (\bibinfo {year} {2019})}\BibitemShut {NoStop}%
\bibitem [{\citenamefont {Yuce}\ and\ \citenamefont {Ramezani}(2019)}]{PhysRevA.100.032102}%
  \BibitemOpen
  \bibfield  {author} {\bibinfo {author} {\bibfnamefont {C.}~\bibnamefont {Yuce}}\ and\ \bibinfo {author} {\bibfnamefont {H.}~\bibnamefont {Ramezani}},\ }\bibfield  {title} {\enquote {\bibinfo {title} {Topological states in a non-hermitian two-dimensional su-schrieffer-heeger model}}, }\href {\doibase 10.1103/PhysRevA.100.032102} {\bibfield  {journal} {\bibinfo  {journal} {Phys. Rev. A}\ }\textbf {\bibinfo {volume} {100}},\ \bibinfo {pages} {032102} (\bibinfo {year} {2019})}\BibitemShut {NoStop}%
\bibitem [{\citenamefont {Chen}\ \emph {et~al.}(2019)\citenamefont {Chen}, \citenamefont {Chen}, \citenamefont {Zhou},\ and\ \citenamefont {Xu}}]{PhysRevB.99.155431}%
  \BibitemOpen
  \bibfield  {author} {\bibinfo {author} {\bibfnamefont {R.}~\bibnamefont {Chen}}, \bibinfo {author} {\bibfnamefont {C.-Z.}\ \bibnamefont {Chen}}, \bibinfo {author} {\bibfnamefont {B.}~\bibnamefont {Zhou}}, \ and\ \bibinfo {author} {\bibfnamefont {D.-H.}\ \bibnamefont {Xu}},\ }\bibfield  {title} {\enquote {\bibinfo {title} {Finite-size effects in non-hermitian topological systems}}, }\href {\doibase 10.1103/PhysRevB.99.155431} {\bibfield  {journal} {\bibinfo  {journal} {Phys. Rev. B}\ }\textbf {\bibinfo {volume} {99}},\ \bibinfo {pages} {155431} (\bibinfo {year} {2019})}\BibitemShut {NoStop}%
\bibitem [{\citenamefont {Yu}\ \emph {et~al.}(2020)\citenamefont {Yu}, \citenamefont {Jiang}, \citenamefont {Quan}, \citenamefont {Wu}, \citenamefont {Chen}, \citenamefont {Zou},\ and\ \citenamefont {Wu}}]{PhysRevB.101.045422}%
  \BibitemOpen
  \bibfield  {author} {\bibinfo {author} {\bibfnamefont {X.-L.}\ \bibnamefont {Yu}}, \bibinfo {author} {\bibfnamefont {L.}~\bibnamefont {Jiang}}, \bibinfo {author} {\bibfnamefont {Y.-M.}\ \bibnamefont {Quan}}, \bibinfo {author} {\bibfnamefont {T.}~\bibnamefont {Wu}}, \bibinfo {author} {\bibfnamefont {Y.}~\bibnamefont {Chen}}, \bibinfo {author} {\bibfnamefont {L.-J.}\ \bibnamefont {Zou}}, \ and\ \bibinfo {author} {\bibfnamefont {J.}~\bibnamefont {Wu}},\ }\bibfield  {title} {\enquote {\bibinfo {title} {Topological phase transitions, majorana modes, and quantum simulation of the su--schrieffer--heeger model with nearest-neighbor interactions}}, }\href {\doibase 10.1103/PhysRevB.101.045422} {\bibfield  {journal} {\bibinfo  {journal} {Phys. Rev. B}\ }\textbf {\bibinfo {volume} {101}},\ \bibinfo {pages} {045422} (\bibinfo {year} {2020})}\BibitemShut {NoStop}%
\bibitem [{\citenamefont {Henriques}\ \emph {et~al.}(2020)\citenamefont {Henriques}, \citenamefont {Rappoport}, \citenamefont {Bludov}, \citenamefont {Vasilevskiy},\ and\ \citenamefont {Peres}}]{PhysRevA.101.043811}%
  \BibitemOpen
  \bibfield  {author} {\bibinfo {author} {\bibfnamefont {J.~C.~G.}\ \bibnamefont {Henriques}}, \bibinfo {author} {\bibfnamefont {T.~G.}\ \bibnamefont {Rappoport}}, \bibinfo {author} {\bibfnamefont {Y.~V.}\ \bibnamefont {Bludov}}, \bibinfo {author} {\bibfnamefont {M.~I.}\ \bibnamefont {Vasilevskiy}}, \ and\ \bibinfo {author} {\bibfnamefont {N.~M.~R.}\ \bibnamefont {Peres}},\ }\bibfield  {title} {\enquote {\bibinfo {title} {Topological photonic tamm states and the su-schrieffer-heeger model}}, }\href {\doibase 10.1103/PhysRevA.101.043811} {\bibfield  {journal} {\bibinfo  {journal} {Phys. Rev. A}\ }\textbf {\bibinfo {volume} {101}},\ \bibinfo {pages} {043811} (\bibinfo {year} {2020})}\BibitemShut {NoStop}%
\bibitem [{\citenamefont {Zeng}\ \emph {et~al.}(2020)\citenamefont {Zeng}, \citenamefont {Yang},\ and\ \citenamefont {Xu}}]{Zeng20prb}%
  \BibitemOpen
  \bibfield  {author} {\bibinfo {author} {\bibfnamefont {Q.-B.}\ \bibnamefont {Zeng}}, \bibinfo {author} {\bibfnamefont {Y.-B.}\ \bibnamefont {Yang}}, \ and\ \bibinfo {author} {\bibfnamefont {Y.}~\bibnamefont {Xu}},\ }\bibfield  {title} {\enquote {\bibinfo {title} {Topological phases in non-hermitian aubry-andr\'e-harper models}}, }\href {\doibase 10.1103/PhysRevB.101.020201} {\bibfield  {journal} {\bibinfo  {journal} {Phys. Rev. B}\ }\textbf {\bibinfo {volume} {101}},\ \bibinfo {pages} {020201} (\bibinfo {year} {2020})}\BibitemShut {NoStop}%
\bibitem [{\citenamefont {Hu}\ and\ \citenamefont {Zhao}(2021)}]{HuH21prl}%
  \BibitemOpen
  \bibfield  {author} {\bibinfo {author} {\bibfnamefont {H.}~\bibnamefont {Hu}}\ and\ \bibinfo {author} {\bibfnamefont {E.}~\bibnamefont {Zhao}},\ }\bibfield  {title} {\enquote {\bibinfo {title} {Knots and non-hermitian bloch bands}}, }\href {\doibase 10.1103/PhysRevLett.126.010401} {\bibfield  {journal} {\bibinfo  {journal} {Phys. Rev. Lett.}\ }\textbf {\bibinfo {volume} {126}},\ \bibinfo {pages} {010401} (\bibinfo {year} {2021})}\BibitemShut {NoStop}%
\bibitem [{\citenamefont {Guo}\ \emph {et~al.}(2021)\citenamefont {Guo}, \citenamefont {Liu}, \citenamefont {Zhao}, \citenamefont {Liu},\ and\ \citenamefont {Chen}}]{GuoCX21prl}%
  \BibitemOpen
  \bibfield  {author} {\bibinfo {author} {\bibfnamefont {C.-X.}\ \bibnamefont {Guo}}, \bibinfo {author} {\bibfnamefont {C.-H.}\ \bibnamefont {Liu}}, \bibinfo {author} {\bibfnamefont {X.-M.}\ \bibnamefont {Zhao}}, \bibinfo {author} {\bibfnamefont {Y.}~\bibnamefont {Liu}}, \ and\ \bibinfo {author} {\bibfnamefont {S.}~\bibnamefont {Chen}},\ }\bibfield  {title} {\enquote {\bibinfo {title} {Exact solution of non-hermitian systems with generalized boundary conditions: Size-dependent boundary effect and fragility of the skin effect}}, }\href {\doibase 10.1103/PhysRevLett.127.116801} {\bibfield  {journal} {\bibinfo  {journal} {Phys. Rev. Lett.}\ }\textbf {\bibinfo {volume} {127}},\ \bibinfo {pages} {116801} (\bibinfo {year} {2021})}\BibitemShut {NoStop}%
\bibitem [{\citenamefont {Li}\ \emph {et~al.}(2021)\citenamefont {Li}, \citenamefont {Zhang},\ and\ \citenamefont {Yi}}]{TianyuLi2021}%
  \BibitemOpen
  \bibfield  {author} {\bibinfo {author} {\bibfnamefont {T.}~\bibnamefont {Li}}, \bibinfo {author} {\bibfnamefont {Y.-S.}\ \bibnamefont {Zhang}}, \ and\ \bibinfo {author} {\bibfnamefont {W.}~\bibnamefont {Yi}},\ }\bibfield  {title} {\enquote {\bibinfo {title} {Two-dimensional quantum walk with non-hermitian skin effects}}, }\href {\doibase 10.1088/0256-307X/38/3/030301} {\bibfield  {journal} {\bibinfo  {journal} {Chinese Physics Letters}\ }\textbf {\bibinfo {volume} {38}},\ \bibinfo {eid} {030301} (\bibinfo {year} {2021})}\BibitemShut {NoStop}%
\bibitem [{\citenamefont {Bergholtz}\ \emph {et~al.}(2021)\citenamefont {Bergholtz}, \citenamefont {Budich},\ and\ \citenamefont {Kunst}}]{Bergholtz21rmp}%
  \BibitemOpen
  \bibfield  {author} {\bibinfo {author} {\bibfnamefont {E.~J.}\ \bibnamefont {Bergholtz}}, \bibinfo {author} {\bibfnamefont {J.~C.}\ \bibnamefont {Budich}}, \ and\ \bibinfo {author} {\bibfnamefont {F.~K.}\ \bibnamefont {Kunst}},\ }\bibfield  {title} {\enquote {\bibinfo {title} {Exceptional topology of non-hermitian systems}}, }\href {\doibase 10.1103/RevModPhys.93.015005} {\bibfield  {journal} {\bibinfo  {journal} {Rev. Mod. Phys.}\ }\textbf {\bibinfo {volume} {93}},\ \bibinfo {pages} {015005} (\bibinfo {year} {2021})}\BibitemShut {NoStop}%
\bibitem [{\citenamefont {Xiujuan~Zhang}\ and\ \citenamefont {Chen}(2022)}]{reviewskineffect}%
  \BibitemOpen
  \bibfield  {author} {\bibinfo {author} {\bibfnamefont {M.-H.~L.}\ \bibnamefont {Xiujuan~Zhang}, \bibfnamefont {Tian~Zhang}}\ and\ \bibinfo {author} {\bibfnamefont {Y.-F.}\ \bibnamefont {Chen}},\ }\bibfield  {title} {\enquote {\bibinfo {title} {A review on non-hermitian skin effect}}, }\href {\doibase 10.1080/23746149.2022.2109431} {\bibfield  {journal} {\bibinfo  {journal} {Advances in Physics: X}\ }\textbf {\bibinfo {volume} {7}},\ \bibinfo {pages} {2109431} (\bibinfo {year} {2022})},\ \Eprint {http://arxiv.org/abs/https://doi.org/10.1080/23746149.2022.2109431} {https://doi.org/10.1080/23746149.2022.2109431} \BibitemShut {NoStop}%
\bibitem [{\citenamefont {Cheng}\ \emph {et~al.}(2022)\citenamefont {Cheng}, \citenamefont {Zhang}, \citenamefont {Lu},\ and\ \citenamefont {Chen}}]{PhysRevB.105.094103}%
  \BibitemOpen
  \bibfield  {author} {\bibinfo {author} {\bibfnamefont {J.}~\bibnamefont {Cheng}}, \bibinfo {author} {\bibfnamefont {X.}~\bibnamefont {Zhang}}, \bibinfo {author} {\bibfnamefont {M.-H.}\ \bibnamefont {Lu}}, \ and\ \bibinfo {author} {\bibfnamefont {Y.-F.}\ \bibnamefont {Chen}},\ }\bibfield  {title} {\enquote {\bibinfo {title} {Competition between band topology and non-hermiticity}}, }\href {\doibase 10.1103/PhysRevB.105.094103} {\bibfield  {journal} {\bibinfo  {journal} {Phys. Rev. B}\ }\textbf {\bibinfo {volume} {105}},\ \bibinfo {pages} {094103} (\bibinfo {year} {2022})}\BibitemShut {NoStop}%
\bibitem [{\citenamefont {Hu}\ \emph {et~al.}(2023)\citenamefont {Hu}, \citenamefont {Fu},\ and\ \citenamefont {Zhang}}]{PhysRevB.108.245114}%
  \BibitemOpen
  \bibfield  {author} {\bibinfo {author} {\bibfnamefont {S.-X.}\ \bibnamefont {Hu}}, \bibinfo {author} {\bibfnamefont {Y.}~\bibnamefont {Fu}}, \ and\ \bibinfo {author} {\bibfnamefont {Y.}~\bibnamefont {Zhang}},\ }\bibfield  {title} {\enquote {\bibinfo {title} {Nontrivial worldline winding in non-hermitian quantum systems}}, }\href {\doibase 10.1103/PhysRevB.108.245114} {\bibfield  {journal} {\bibinfo  {journal} {Phys. Rev. B}\ }\textbf {\bibinfo {volume} {108}},\ \bibinfo {pages} {245114} (\bibinfo {year} {2023})}\BibitemShut {NoStop}%
\bibitem [{\citenamefont {Yoshida}\ \emph {et~al.}(2024)\citenamefont {Yoshida}, \citenamefont {Zhang}, \citenamefont {Neupert},\ and\ \citenamefont {Kawakami}}]{PhysRevLett.133.076502}%
  \BibitemOpen
  \bibfield  {author} {\bibinfo {author} {\bibfnamefont {T.}~\bibnamefont {Yoshida}}, \bibinfo {author} {\bibfnamefont {S.-B.}\ \bibnamefont {Zhang}}, \bibinfo {author} {\bibfnamefont {T.}~\bibnamefont {Neupert}}, \ and\ \bibinfo {author} {\bibfnamefont {N.}~\bibnamefont {Kawakami}},\ }\bibfield  {title} {\enquote {\bibinfo {title} {Non-hermitian mott skin effect}}, }\href {\doibase 10.1103/PhysRevLett.133.076502} {\bibfield  {journal} {\bibinfo  {journal} {Phys. Rev. Lett.}\ }\textbf {\bibinfo {volume} {133}},\ \bibinfo {pages} {076502} (\bibinfo {year} {2024})}\BibitemShut {NoStop}%
\bibitem [{\citenamefont {Wang}\ \emph {et~al.}(2024)\citenamefont {Wang}, \citenamefont {Song},\ and\ \citenamefont {Wang}}]{PRXwangzhong}%
  \BibitemOpen
  \bibfield  {author} {\bibinfo {author} {\bibfnamefont {H.-Y.}\ \bibnamefont {Wang}}, \bibinfo {author} {\bibfnamefont {F.}~\bibnamefont {Song}}, \ and\ \bibinfo {author} {\bibfnamefont {Z.}~\bibnamefont {Wang}},\ }\bibfield  {title} {\enquote {\bibinfo {title} {Amoeba formulation of non-bloch band theory in arbitrary dimensions}}, }\href {\doibase 10.1103/PhysRevX.14.021011} {\bibfield  {journal} {\bibinfo  {journal} {Phys. Rev. X}\ }\textbf {\bibinfo {volume} {14}},\ \bibinfo {pages} {021011} (\bibinfo {year} {2024})}\BibitemShut {NoStop}%
\bibitem [{\citenamefont {Hou}\ \emph {et~al.}(2024)\citenamefont {Hou}, \citenamefont {Tang}, \citenamefont {Xu},\ and\ \citenamefont {Lin}}]{WanerHou2024}%
  \BibitemOpen
  \bibfield  {author} {\bibinfo {author} {\bibfnamefont {W.}~\bibnamefont {Hou}}, \bibinfo {author} {\bibfnamefont {H.}~\bibnamefont {Tang}}, \bibinfo {author} {\bibfnamefont {Q.}~\bibnamefont {Xu}}, \ and\ \bibinfo {author} {\bibfnamefont {Y.}~\bibnamefont {Lin}},\ }\bibfield  {title} {\enquote {\bibinfo {title} {Experimental proposal on non-hermitian skin effect by two-dimensional quantum walk with a single trapped ion}}, }\href {\doibase 10.1088/0256-307X/41/4/040301} {\bibfield  {journal} {\bibinfo  {journal} {Chinese Physics Letters}\ }\textbf {\bibinfo {volume} {41}},\ \bibinfo {eid} {040301} (\bibinfo {year} {2024})}\BibitemShut {NoStop}%
\bibitem [{\citenamefont {Yokomizo}\ and\ \citenamefont {Murakami}(2019)}]{PhysRevLett.123.066404}%
  \BibitemOpen
  \bibfield  {author} {\bibinfo {author} {\bibfnamefont {K.}~\bibnamefont {Yokomizo}}\ and\ \bibinfo {author} {\bibfnamefont {S.}~\bibnamefont {Murakami}},\ }\bibfield  {title} {\enquote {\bibinfo {title} {Non-bloch band theory of non-hermitian systems}}, }\href {\doibase 10.1103/PhysRevLett.123.066404} {\bibfield  {journal} {\bibinfo  {journal} {Phys. Rev. Lett.}\ }\textbf {\bibinfo {volume} {123}},\ \bibinfo {pages} {066404} (\bibinfo {year} {2019})}\BibitemShut {NoStop}%
\bibitem [{\citenamefont {Lee}\ \emph {et~al.}(2019)\citenamefont {Lee}, \citenamefont {Li},\ and\ \citenamefont {Gong}}]{PhysRevLett.123.016805}%
  \BibitemOpen
  \bibfield  {author} {\bibinfo {author} {\bibfnamefont {C.~H.}\ \bibnamefont {Lee}}, \bibinfo {author} {\bibfnamefont {L.}~\bibnamefont {Li}}, \ and\ \bibinfo {author} {\bibfnamefont {J.}~\bibnamefont {Gong}},\ }\bibfield  {title} {\enquote {\bibinfo {title} {{Hybrid Higher}-{Order Skin}-{Topological Modes} in {Nonreciprocal Systems}}}, }\href {\doibase 10.1103/PhysRevLett.123.016805} {\bibfield  {journal} {\bibinfo  {journal} {Phys. Rev. Lett.}\ }\textbf {\bibinfo {volume} {123}},\ \bibinfo {pages} {016805} (\bibinfo {year} {2019})}\BibitemShut {NoStop}%
\bibitem [{\citenamefont {Kawabata}\ \emph {et~al.}(2020)\citenamefont {Kawabata}, \citenamefont {Sato},\ and\ \citenamefont {Shiozaki}}]{PhysRevB.102.205118}%
  \BibitemOpen
  \bibfield  {author} {\bibinfo {author} {\bibfnamefont {K.}~\bibnamefont {Kawabata}}, \bibinfo {author} {\bibfnamefont {M.}~\bibnamefont {Sato}}, \ and\ \bibinfo {author} {\bibfnamefont {K.}~\bibnamefont {Shiozaki}},\ }\bibfield  {title} {\enquote {\bibinfo {title} {Higher-order non-{Hermitian} skin effect}}, }\href {\doibase 10.1103/PhysRevB.102.205118} {\bibfield  {journal} {\bibinfo  {journal} {Phys. Rev. B}\ }\textbf {\bibinfo {volume} {102}},\ \bibinfo {pages} {205118} (\bibinfo {year} {2020})}\BibitemShut {NoStop}%
\bibitem [{\citenamefont {Okugawa}\ \emph {et~al.}(2020)\citenamefont {Okugawa}, \citenamefont {Takahashi},\ and\ \citenamefont {Yokomizo}}]{PhysRevB.102.241202}%
  \BibitemOpen
  \bibfield  {author} {\bibinfo {author} {\bibfnamefont {R.}~\bibnamefont {Okugawa}}, \bibinfo {author} {\bibfnamefont {R.}~\bibnamefont {Takahashi}}, \ and\ \bibinfo {author} {\bibfnamefont {K.}~\bibnamefont {Yokomizo}},\ }\bibfield  {title} {\enquote {\bibinfo {title} {Second-order topological non-{Hermitian} skin effects}}, }\href {\doibase 10.1103/PhysRevB.102.241202} {\bibfield  {journal} {\bibinfo  {journal} {Phys. Rev. B}\ }\textbf {\bibinfo {volume} {102}},\ \bibinfo {pages} {241202} (\bibinfo {year} {2020})}\BibitemShut {NoStop}%
\bibitem [{\citenamefont {Fu}\ \emph {et~al.}(2021)\citenamefont {Fu}, \citenamefont {Hu},\ and\ \citenamefont {Wan}}]{FuY21prb}%
  \BibitemOpen
  \bibfield  {author} {\bibinfo {author} {\bibfnamefont {Y.}~\bibnamefont {Fu}}, \bibinfo {author} {\bibfnamefont {J.}~\bibnamefont {Hu}}, \ and\ \bibinfo {author} {\bibfnamefont {S.}~\bibnamefont {Wan}},\ }\bibfield  {title} {\enquote {\bibinfo {title} {Non-hermitian second-order skin and topological modes}}, }\href {\doibase 10.1103/PhysRevB.103.045420} {\bibfield  {journal} {\bibinfo  {journal} {Phys. Rev. B}\ }\textbf {\bibinfo {volume} {103}},\ \bibinfo {pages} {045420} (\bibinfo {year} {2021})}\BibitemShut {NoStop}%
\bibitem [{\citenamefont {Zhang}\ \emph {et~al.}(2021)\citenamefont {Zhang}, \citenamefont {Tian}, \citenamefont {Jiang}, \citenamefont {Lu},\ and\ \citenamefont {Chen}}]{Zhangx21nc}%
  \BibitemOpen
  \bibfield  {author} {\bibinfo {author} {\bibfnamefont {X.}~\bibnamefont {Zhang}}, \bibinfo {author} {\bibfnamefont {Y.}~\bibnamefont {Tian}}, \bibinfo {author} {\bibfnamefont {J.-H.}\ \bibnamefont {Jiang}}, \bibinfo {author} {\bibfnamefont {M.-H.}\ \bibnamefont {Lu}}, \ and\ \bibinfo {author} {\bibfnamefont {Y.-F.}\ \bibnamefont {Chen}},\ }\bibfield  {title} {\enquote {\bibinfo {title} {Observation of higher-order non-hermitian skin effect}}, }\href {\doibase 10.1038/s41467-021-25716-y} {\bibfield  {journal} {\bibinfo  {journal} {Nature Communications}\ }\textbf {\bibinfo {volume} {12}},\ \bibinfo {pages} {5377} (\bibinfo {year} {2021})}\BibitemShut {NoStop}%
\bibitem [{\citenamefont {Li}\ \emph {et~al.}(2022)\citenamefont {Li}, \citenamefont {Liang}, \citenamefont {Wang}, \citenamefont {Lu},\ and\ \citenamefont {Liu}}]{PhysRevLett.128.223903}%
  \BibitemOpen
  \bibfield  {author} {\bibinfo {author} {\bibfnamefont {Y.}~\bibnamefont {Li}}, \bibinfo {author} {\bibfnamefont {C.}~\bibnamefont {Liang}}, \bibinfo {author} {\bibfnamefont {C.}~\bibnamefont {Wang}}, \bibinfo {author} {\bibfnamefont {C.}~\bibnamefont {Lu}}, \ and\ \bibinfo {author} {\bibfnamefont {Y.-C.}\ \bibnamefont {Liu}},\ }\bibfield  {title} {\enquote {\bibinfo {title} {Gain-{Loss}-{Induced Hybrid Skin}-{Topological Effect}}}, }\href {\doibase 10.1103/PhysRevLett.128.223903} {\bibfield  {journal} {\bibinfo  {journal} {Phys. Rev. Lett.}\ }\textbf {\bibinfo {volume} {128}},\ \bibinfo {pages} {223903} (\bibinfo {year} {2022})}\BibitemShut {NoStop}%
\bibitem [{\citenamefont {Zhu}\ and\ \citenamefont {Gong}(2022)}]{PhysRevB.106.035425}%
  \BibitemOpen
  \bibfield  {author} {\bibinfo {author} {\bibfnamefont {W.}~\bibnamefont {Zhu}}\ and\ \bibinfo {author} {\bibfnamefont {J.}~\bibnamefont {Gong}},\ }\bibfield  {title} {\enquote {\bibinfo {title} {Hybrid skin-topological modes without asymmetric couplings}}, }\href {\doibase 10.1103/PhysRevB.106.035425} {\bibfield  {journal} {\bibinfo  {journal} {Phys. Rev. B}\ }\textbf {\bibinfo {volume} {106}},\ \bibinfo {pages} {035425} (\bibinfo {year} {2022})}\BibitemShut {NoStop}%
\bibitem [{\citenamefont {Li}\ \emph {et~al.}(2023)\citenamefont {Li}, \citenamefont {Trauzettel}, \citenamefont {Neupert},\ and\ \citenamefont {Zhang}}]{PhysRevLett.131.116601}%
  \BibitemOpen
  \bibfield  {author} {\bibinfo {author} {\bibfnamefont {C.-A.}\ \bibnamefont {Li}}, \bibinfo {author} {\bibfnamefont {B.}~\bibnamefont {Trauzettel}}, \bibinfo {author} {\bibfnamefont {T.}~\bibnamefont {Neupert}}, \ and\ \bibinfo {author} {\bibfnamefont {S.-B.}\ \bibnamefont {Zhang}},\ }\bibfield  {title} {\enquote {\bibinfo {title} {Enhancement of {Second}-{Order} {Non}-{Hermitian} {Skin Effect} by {Magnetic Fields}}}, }\href {\doibase 10.1103/PhysRevLett.131.116601} {\bibfield  {journal} {\bibinfo  {journal} {Phys. Rev. Lett.}\ }\textbf {\bibinfo {volume} {131}},\ \bibinfo {pages} {116601} (\bibinfo {year} {2023})}\BibitemShut {NoStop}%
\bibitem [{\citenamefont {Sun}\ \emph {et~al.}(2023)\citenamefont {Sun}, \citenamefont {Li}, \citenamefont {Feng},\ and\ \citenamefont {Guo}}]{PhysRevB.108.075122}%
  \BibitemOpen
  \bibfield  {author} {\bibinfo {author} {\bibfnamefont {J.}~\bibnamefont {Sun}}, \bibinfo {author} {\bibfnamefont {C.-A.}\ \bibnamefont {Li}}, \bibinfo {author} {\bibfnamefont {S.}~\bibnamefont {Feng}}, \ and\ \bibinfo {author} {\bibfnamefont {H.}~\bibnamefont {Guo}},\ }\bibfield  {title} {\enquote {\bibinfo {title} {Hybrid higher-order skin-topological effect in hyperbolic lattices}}, }\href {\doibase 10.1103/PhysRevB.108.075122} {\bibfield  {journal} {\bibinfo  {journal} {Phys. Rev. B}\ }\textbf {\bibinfo {volume} {108}},\ \bibinfo {pages} {075122} (\bibinfo {year} {2023})}\BibitemShut {NoStop}%
\bibitem [{\citenamefont {Zhang}\ \emph {et~al.}(2022)\citenamefont {Zhang}, \citenamefont {Yang},\ and\ \citenamefont {Fang}}]{Zhang2022}%
  \BibitemOpen
  \bibfield  {author} {\bibinfo {author} {\bibfnamefont {K.}~\bibnamefont {Zhang}}, \bibinfo {author} {\bibfnamefont {Z.}~\bibnamefont {Yang}}, \ and\ \bibinfo {author} {\bibfnamefont {C.}~\bibnamefont {Fang}},\ }\bibfield  {title} {\enquote {\bibinfo {title} {Universal non-hermitian skin effect in two and higher dimensions}}, }\href {\doibase 10.1038/s41467-022-30161-6} {\bibfield  {journal} {\bibinfo  {journal} {Nature Communications}\ }\textbf {\bibinfo {volume} {13}},\ \bibinfo {pages} {2496} (\bibinfo {year} {2022})}\BibitemShut {NoStop}%
\bibitem [{\citenamefont {Sun}\ \emph {et~al.}(2024)\citenamefont {Sun}, \citenamefont {Li}, \citenamefont {Guo}, \citenamefont {Zhang}, \citenamefont {Feng}, \citenamefont {Zhang}, \citenamefont {Guo},\ and\ \citenamefont {Trauzettel}}]{Sun24NHQF}%
  \BibitemOpen
  \bibfield  {author} {\bibinfo {author} {\bibfnamefont {J.}~\bibnamefont {Sun}}, \bibinfo {author} {\bibfnamefont {C.-A.}\ \bibnamefont {Li}}, \bibinfo {author} {\bibfnamefont {Q.}~\bibnamefont {Guo}}, \bibinfo {author} {\bibfnamefont {W.}~\bibnamefont {Zhang}}, \bibinfo {author} {\bibfnamefont {S.}~\bibnamefont {Feng}}, \bibinfo {author} {\bibfnamefont {X.}~\bibnamefont {Zhang}}, \bibinfo {author} {\bibfnamefont {H.}~\bibnamefont {Guo}}, \ and\ \bibinfo {author} {\bibfnamefont {B.}~\bibnamefont {Trauzettel}},\ }\href {https://arxiv.org/abs/2408.07355} {\enquote {\bibinfo {title} {Non-hermitian quantum fractals}}, } (\bibinfo {year} {2024}),\ \Eprint {http://arxiv.org/abs/2408.07355} {arXiv:2408.07355 [cond-mat.mes-hall]} \BibitemShut {NoStop}%
\bibitem [{\citenamefont {Manna}\ and\ \citenamefont {Roy}(2023)}]{innerskin}%
  \BibitemOpen
  \bibfield  {author} {\bibinfo {author} {\bibfnamefont {S.}~\bibnamefont {Manna}}\ and\ \bibinfo {author} {\bibfnamefont {B.}~\bibnamefont {Roy}},\ }\bibfield  {title} {\enquote {\bibinfo {title} {Inner skin effects on non-hermitian topological fractals}}, }\href {\doibase 10.1038/s42005-023-01130-2} {\bibfield  {journal} {\bibinfo  {journal} {Communications Physics}\ }\textbf {\bibinfo {volume} {6}},\ \bibinfo {pages} {10} (\bibinfo {year} {2023})}\BibitemShut {NoStop}%
\bibitem [{\citenamefont {Eliashvili}\ \emph {et~al.}(2017)\citenamefont {Eliashvili}, \citenamefont {Kereselidze}, \citenamefont {Tsitsishvili},\ and\ \citenamefont {Tsitsishvili}}]{jps2017}%
  \BibitemOpen
  \bibfield  {author} {\bibinfo {author} {\bibfnamefont {M.}~\bibnamefont {Eliashvili}}, \bibinfo {author} {\bibfnamefont {D.}~\bibnamefont {Kereselidze}}, \bibinfo {author} {\bibfnamefont {G.}~\bibnamefont {Tsitsishvili}}, \ and\ \bibinfo {author} {\bibfnamefont {M.}~\bibnamefont {Tsitsishvili}},\ }\bibfield  {title} {\enquote {\bibinfo {title} {{Edge States} of a {Periodic Chain} with {Four}-{Band Energy Spectrum}}}, }\href {\doibase 10.7566/JPSJ.86.074712} {\bibfield  {journal} {\bibinfo  {journal} {Journal of the Physical Society of Japan}\ }\textbf {\bibinfo {volume} {86}},\ \bibinfo {pages} {074712} (\bibinfo {year} {2017})},\ \Eprint {http://arxiv.org/abs/https://doi.org/10.7566/JPSJ.86.074712} {https://doi.org/10.7566/JPSJ.86.074712} \BibitemShut {NoStop}%
\bibitem [{\citenamefont {Lee}\ \emph {et~al.}(2018)\citenamefont {Lee}, \citenamefont {Imhof}, \citenamefont {Berger}, \citenamefont {Bayer}, \citenamefont {Brehm}, \citenamefont {Molenkamp}, \citenamefont {Kiessling},\ and\ \citenamefont {Thomale}}]{Lee2018}%
  \BibitemOpen
  \bibfield  {author} {\bibinfo {author} {\bibfnamefont {C.~H.}\ \bibnamefont {Lee}}, \bibinfo {author} {\bibfnamefont {S.}~\bibnamefont {Imhof}}, \bibinfo {author} {\bibfnamefont {C.}~\bibnamefont {Berger}}, \bibinfo {author} {\bibfnamefont {F.}~\bibnamefont {Bayer}}, \bibinfo {author} {\bibfnamefont {J.}~\bibnamefont {Brehm}}, \bibinfo {author} {\bibfnamefont {L.~W.}\ \bibnamefont {Molenkamp}}, \bibinfo {author} {\bibfnamefont {T.}~\bibnamefont {Kiessling}}, \ and\ \bibinfo {author} {\bibfnamefont {R.}~\bibnamefont {Thomale}},\ }\bibfield  {title} {\enquote {\bibinfo {title} {Topolectrical circuits}}, }\href {\doibase 10.1038/s42005-018-0035-2} {\bibfield  {journal} {\bibinfo  {journal} {Communications Physics}\ }\textbf {\bibinfo {volume} {1}},\ \bibinfo {pages} {39} (\bibinfo {year} {2018})}\BibitemShut {NoStop}%
\bibitem [{\citenamefont {Dong}\ \emph {et~al.}(2021)\citenamefont {Dong}, \citenamefont {Juri\ifmmode \check{c}\else \v{c}\fi{}i\ifmmode~\acute{c}\else \'{c}\fi{}},\ and\ \citenamefont {Roy}}]{PhysRevResearch.3.023056}%
  \BibitemOpen
  \bibfield  {author} {\bibinfo {author} {\bibfnamefont {J.}~\bibnamefont {Dong}}, \bibinfo {author} {\bibfnamefont {V.}~\bibnamefont {Juri\ifmmode \check{c}\else \v{c}\fi{}i\ifmmode~\acute{c}\else \'{c}\fi{}}}, \ and\ \bibinfo {author} {\bibfnamefont {B.}~\bibnamefont {Roy}},\ }\bibfield  {title} {\enquote {\bibinfo {title} {Topolectric circuits: Theory and construction}}, }\href {\doibase 10.1103/PhysRevResearch.3.023056} {\bibfield  {journal} {\bibinfo  {journal} {Phys. Rev. Res.}\ }\textbf {\bibinfo {volume} {3}},\ \bibinfo {pages} {023056} (\bibinfo {year} {2021})}\BibitemShut {NoStop}%
\bibitem [{\citenamefont {Xu}\ \emph {et~al.}(2021)\citenamefont {Xu}, \citenamefont {Zhang}, \citenamefont {Luo}, \citenamefont {Yu}, \citenamefont {Li},\ and\ \citenamefont {Zhang}}]{PhysRevB.103.125411}%
  \BibitemOpen
  \bibfield  {author} {\bibinfo {author} {\bibfnamefont {K.}~\bibnamefont {Xu}}, \bibinfo {author} {\bibfnamefont {X.}~\bibnamefont {Zhang}}, \bibinfo {author} {\bibfnamefont {K.}~\bibnamefont {Luo}}, \bibinfo {author} {\bibfnamefont {R.}~\bibnamefont {Yu}}, \bibinfo {author} {\bibfnamefont {D.}~\bibnamefont {Li}}, \ and\ \bibinfo {author} {\bibfnamefont {H.}~\bibnamefont {Zhang}},\ }\bibfield  {title} {\enquote {\bibinfo {title} {Coexistence of topological edge states and skin effects in the non-hermitian su-schrieffer-heeger model with long-range nonreciprocal hopping in topoelectric realizations}}, }\href {\doibase 10.1103/PhysRevB.103.125411} {\bibfield  {journal} {\bibinfo  {journal} {Phys. Rev. B}\ }\textbf {\bibinfo {volume} {103}},\ \bibinfo {pages} {125411} (\bibinfo {year} {2021})}\BibitemShut {NoStop}%
\bibitem [{\citenamefont {Rafi-Ul-Islam}\ \emph {et~al.}(2021)\citenamefont {Rafi-Ul-Islam}, \citenamefont {Siu},\ and\ \citenamefont {Jalil}}]{PhysRevB.103.035420}%
  \BibitemOpen
  \bibfield  {author} {\bibinfo {author} {\bibfnamefont {S.~M.}\ \bibnamefont {Rafi-Ul-Islam}}, \bibinfo {author} {\bibfnamefont {Z.~B.}\ \bibnamefont {Siu}}, \ and\ \bibinfo {author} {\bibfnamefont {M.~B.~A.}\ \bibnamefont {Jalil}},\ }\bibfield  {title} {\enquote {\bibinfo {title} {Topological phases with higher winding numbers in nonreciprocal one-dimensional topolectrical circuits}}, }\href {\doibase 10.1103/PhysRevB.103.035420} {\bibfield  {journal} {\bibinfo  {journal} {Phys. Rev. B}\ }\textbf {\bibinfo {volume} {103}},\ \bibinfo {pages} {035420} (\bibinfo {year} {2021})}\BibitemShut {NoStop}%
\bibitem [{\citenamefont {Liu}\ \emph {et~al.}(2022)\citenamefont {Liu}, \citenamefont {Cao}, \citenamefont {Chen}, \citenamefont {Wang}, \citenamefont {Yang},\ and\ \citenamefont {Zhang}}]{Liu2022}%
  \BibitemOpen
  \bibfield  {author} {\bibinfo {author} {\bibfnamefont {Y.}~\bibnamefont {Liu}}, \bibinfo {author} {\bibfnamefont {W.}~\bibnamefont {Cao}}, \bibinfo {author} {\bibfnamefont {W.}~\bibnamefont {Chen}}, \bibinfo {author} {\bibfnamefont {H.}~\bibnamefont {Wang}}, \bibinfo {author} {\bibfnamefont {L.}~\bibnamefont {Yang}}, \ and\ \bibinfo {author} {\bibfnamefont {X.}~\bibnamefont {Zhang}},\ }\bibfield  {title} {\enquote {\bibinfo {title} {Fully integrated topological electronics}}, }\href {\doibase 10.1038/s41598-022-17010-8} {\bibfield  {journal} {\bibinfo  {journal} {Scientific Reports}\ }\textbf {\bibinfo {volume} {12}},\ \bibinfo {pages} {13410} (\bibinfo {year} {2022})}\BibitemShut {NoStop}%
\bibitem [{\citenamefont {Rafi-Ul-Islam}\ \emph {et~al.}(2023)\citenamefont {Rafi-Ul-Islam}, \citenamefont {Siu}, \citenamefont {Sahin},\ and\ \citenamefont {Jalil}}]{PhysRevResearch.5.013107}%
  \BibitemOpen
  \bibfield  {author} {\bibinfo {author} {\bibfnamefont {S.~M.}\ \bibnamefont {Rafi-Ul-Islam}}, \bibinfo {author} {\bibfnamefont {Z.~B.}\ \bibnamefont {Siu}}, \bibinfo {author} {\bibfnamefont {H.}~\bibnamefont {Sahin}}, \ and\ \bibinfo {author} {\bibfnamefont {M.~B.~A.}\ \bibnamefont {Jalil}},\ }\bibfield  {title} {\enquote {\bibinfo {title} {Valley hall effect and kink states in topolectrical circuits}}, }\href {\doibase 10.1103/PhysRevResearch.5.013107} {\bibfield  {journal} {\bibinfo  {journal} {Phys. Rev. Res.}\ }\textbf {\bibinfo {volume} {5}},\ \bibinfo {pages} {013107} (\bibinfo {year} {2023})}\BibitemShut {NoStop}%
\bibitem [{\citenamefont {Hamanaka}\ \emph {et~al.}(2024)\citenamefont {Hamanaka}, \citenamefont {Iliasov}, \citenamefont {Neupert}, \citenamefont {Bzdušek},\ and\ \citenamefont {Yoshida}}]{multifractal}%
  \BibitemOpen
  \bibfield  {author} {\bibinfo {author} {\bibfnamefont {S.}~\bibnamefont {Hamanaka}}, \bibinfo {author} {\bibfnamefont {A.~A.}\ \bibnamefont {Iliasov}}, \bibinfo {author} {\bibfnamefont {T.}~\bibnamefont {Neupert}}, \bibinfo {author} {\bibfnamefont {T.}~\bibnamefont {Bzdušek}}, \ and\ \bibinfo {author} {\bibfnamefont {T.}~\bibnamefont {Yoshida}},\ }\href {https://arxiv.org/abs/2408.11024} {\enquote {\bibinfo {title} {Multifractal statistics of non-hermitian skin effect on the cayley tree}}, } (\bibinfo {year} {2024}),\ \Eprint {http://arxiv.org/abs/2408.11024} {arXiv:2408.11024 [cond-mat.mes-hall]} \BibitemShut {NoStop}%
\bibitem [{\citenamefont {Hatano}\ \emph {et~al.}(2024)\citenamefont {Hatano}, \citenamefont {Katsura},\ and\ \citenamefont {Kawabata}}]{quantumtransportbethe}%
  \BibitemOpen
  \bibfield  {author} {\bibinfo {author} {\bibfnamefont {N.}~\bibnamefont {Hatano}}, \bibinfo {author} {\bibfnamefont {H.}~\bibnamefont {Katsura}}, \ and\ \bibinfo {author} {\bibfnamefont {K.}~\bibnamefont {Kawabata}},\ }\href {https://arxiv.org/abs/2409.01873} {\enquote {\bibinfo {title} {Quantum transport on bethe lattices with non-hermitian sources and a drain}}, } (\bibinfo {year} {2024}),\ \Eprint {http://arxiv.org/abs/2409.01873} {arXiv:2409.01873 [quant-ph]} \BibitemShut {NoStop}%
\end{thebibliography}%

\end{document}